\documentclass[amsmath,amssymb,amsbsy,reprint,prxq,preprintnumbers,showpacs,superscriptaddress]{revtex4-2}
\usepackage{graphicx,color}
\usepackage{dcolumn}
\usepackage{bm}
\usepackage{braket}
\usepackage{mathtools}
\usepackage{ulem}
\usepackage[breaklinks,colorlinks=true,linkcolor=blue,urlcolor=blue,citecolor=blue]{hyperref}
\usepackage{times}
\usepackage{physics}
\usepackage{latexsym}
\usepackage{amsmath, amssymb}
\usepackage{mathtools}
\usepackage{multirow}
\usepackage{arydshln}

\newcommand{\be}{\begin{eqnarray}}
\newcommand{\ee}{\end{eqnarray}}

\renewcommand{\theequation}{\arabic{equation}}

\begin{document}

\title{Decohered color code and emerging mixed toric code
by anyon proliferation: Topological entanglement negativity perspective}
\date{\today}
\author{Keisuke Kataoka}
\thanks{These authors equally contributed}
\affiliation{Department of General Education, Faculty of Science and Technology, Meijo University, Nagoya 468-8502, Japan}
\author{Yoshihito Kuno}
\thanks{These authors equally contributed}
\affiliation{Graduate School of Engineering Science, Akita University, Akita 010-8502, Japan}
\author{Takahiro Orito}
\thanks{These authors equally contributed}
\affiliation{Department of Physics, College of Humanities and Sciences, Nihon University, Sakurajosui, Setagaya, Tokyo 156-8550, Japan}
\author{Ikuo Ichinose} 
\thanks{A professor emeritus}
\affiliation{Department of Applied Physics, Nagoya Institute of Technology, Nagoya 466-8555, Japan}


\begin{abstract} 
This work clarifies how color code under decoherence becomes an intrinsic mixed state with
topological order (imTO), which has no counterparts in the pure ground states of local gaped Hamiltonian. 
We find that for the specific decoherence realized with $XX$-type operators on red links of 
a honeycomb lattice,
the imTO inherits the half of the topological properties of the color code such as
the anyon data, topological entanglement and logical operators.
This work focuses on both pure and mixed stabilizer topological codes. 
In this setup, a gauging out procedure, which is proposed for understanding the mixed states from the
viewpoint of the subsystem code, plays an important pole in identifying the decohered mixed state. 
The properties of the imTO of the decohered color code is clarified from the set of the stabilizer 
generators obtained via the gauging procedure. 
This observation implies that the topological order of the emergent mixed state is close to that of the TC. 
In particular, this imTO can be characterized by topological entanglement negativity (TEN), 
which is a genuine quantity measuring the universal topological nature. 
To investigate the color code under decoherence in detail, we employ efficient numerical algorithm of 
the stabilizer formalism for the target system to observe the entanglement negativity. 
We find that while in the pure color code, the TEN is $2\ln 2$ as expected, in the maximal decohered state, 
the TEN is $\ln 2$, indicating the appearance of a single TC.
Following up the above observation, we extensively study intermediate mixed states between 
the two topological states by controlling the strength of the decoherence, and find that the TEN and its 
variance exhibit interesting behavior,  that is, the TEN changes from $2\ln 2$ to $\ln 2$ rather smoothly 
but its variance has a large and system-size independent peak.
Furthermore, we perform careful numerical study on the negativity and find that it exhibits 
a specific scaling for subsystems that are commensurate with the triangular lattice on which
the emergent TC is defined, whereas the negativity of the other subsystems does not.
This fact obviously indicates that the negativity is a good measure for an appearance
of the TC as a result of the $XX$-type decoherence.
To our best knowledge, no previous works have observed this kind of behavior of the TEN
and the negativity, and therefore, these findings shall shed light on deep understanding on the universal 
characterization of the emerging topological order of mixed states.
\end{abstract}


\maketitle

\section{Introduction}
Promising quantum memory exploits non-trivial properties of quantum many body systems. 
In particular, topological orders, first discovered in quantum Hall phenomena~\cite{Wen_2017},
play a significant role for storing and manipulating quantum information~\cite{Wen_text}. 
That is, the topological phase can be used as quantum devise and a base of quantum 
computations~\cite{Nayak_rev}. 
As the first step for the application of the topological order to quantum information devices, 
toric code (TC) was proposed by Kitaev~\cite{Kitaev_1997,kitaev2003}. 
Just after the proposal, the TC has been extensively studied as a prototypical topological stabilizer code~\cite{gottesman1998,RevModPhys.87.307,fujii2015}. 
Moreover, there exists  important subject, how robust the TC is as a quantum memory. 
In early days, the properties of memory have been extensively investigated such as ~\cite{dennis2002} 
by mapping the TC into statistical mechanical models and, then through lattice gauge theoretical 
viewpoints~\cite{arakawa2004,OHNO_2004462}. 
Subsequently, the robustness against some idealized noise or perturbations, namely error threshold, has been studied in~\cite{dennis2002,Fujii2012}.
Then, the TC provides a fault-tolerant quantum memory where the error recovery process exists 
such as minimum weight perfect matching \cite{dennis2002}. 
It also stems from the gauge-theoretical properties and emergent topological orders \cite{Wen_text}.

From the condensed matter viewpoint, the Hamiltonian of the TC is local, gapped and well isolated 
from the environment. 
The ground state degeneracy depends on the spatial topology~\cite{Wen_2017,Wen_text}.
As the TC is a stabilizer system, the ground states of its Hamiltonian are obtained exactly, and 
then excited states are described by objects named anyons having nontrivial mutual braiding.
Operators creating a pair of anyons are Wilson and 't Hooft strings \cite{PhysRevD.10.2445,thooft1978}, 
which play an important role in the gauge theory classifying its phase diagram. 
Furthermore, the loops of anyons \cite{Kitaev2006} work as logical operators of quantum memory
exploiting topological orders.

After the proposal of the TC, some extension has been proposed. 
One of the most important extensions is color code~ \cite{Bombin2006,Bombin2012,Lidar_Brun_2013,Bombin2015}. 
This code has an additional degrees of freedom, namely ``color''. 
Compared to the TC, this freedom gives rich structure to topological nature beyond the TC, 
allowing us to various applications such as the implementation of quantum teleportation, dense coding 
and computation of magic states, etc.~\cite{Bombin2006,Bombin_2012NJP,Bombin2012}. 
In this work, we focus on the color code.

In recent years, there has been growing interest in the study of the TC and other stabilizer codes 
from the perspective of quantum measurement and noise described as quantum channels.
For example, quantum measurements directly create the stabilizer codes from 
product states \cite{Lavasani2021,PhysRevResearch.6.L042063,Sriram_2023,KOI2024_TC_MoC,OKI2024_TC_MoC,Botzung_2025,Kataoka_2026}. 
The origin of the idea comes from the measurement base quantum computing (MBQC)~\cite{Raussendorf_2001,Raussendorf_2003,Briegel2009}. 
Starting with symmetry protected topological states in 1D or 2D systems or topological ordered states such
as TC, suitable measurements applying to some subsystems create long-range entangled states, 
non-Abelian topological order, and reduced symmetry-protected topological states \cite{verresen_2022,Tsung-Cheng_2023,Tantivasadakarn_2024,KOI_PRB_2024}. 
The generation of nontrivial quantum states through quantum measurements and decoherences,
rather than unitary dynamics, has recently became an active topic of research at the interface 
between quantum many-body physics and quantum information science.

In most of studies given so far, the color code and related stabilizer codes are set on isolated situation. 
However in the real world, environment and noise give significant effects on the codes. 
Under these effects, how topological order such as the color code is robust or changes is one of 
the attracted issue in the quantum information and condensed matter physics communities. 
The TC under decoherence described by quantum channels \cite{Nielsen2011} has been studied in \cite{Fan2024,Sang_PRX_2024,negari2025,Wang_2025,Sohal2025,Sang_PRL_2025,Ellison_PRXQuantum_2025,KOH2025_v1}. 
In these previous works, the pure topologically-ordered state turns into a mixed state. In particular, in \cite{Fan2024}, the original topological order of the TC remains even under local decoherence 
channels up to a threshold of the strength of the decoherence, that is, the topological order remains even in
mixed situation. 
Furthermore, interestingly-enough, the decoherence can induce a non-trivial topological order of mixed 
states called intrinsic mixed state topological order (imTO) proposed in \cite{Wang_2025,Sohal2025,Ellison_PRXQuantum_2025,hsin2025}. 
The imTO is peculiar in a sense that it occurs only mixed states and cannot appear as any
ground states of local and gaped Hamiltonian in $(2+1)$-D 
system\cite{Wang_2025,Ellison_PRXQuantum_2025}. 
As pointed out in Refs. \cite{Sohal2025,Ellison_PRXQuantum_2025}, the imTO possesses some 
transparent anyons. 
Existence of the transparent anyon in the mixed state is related to the emergence of the strong 1-from 
symmetry of the loop operator of the anyon.

There remains an important issue how the imTO is universally characterized. 
In fact, topological entanglement entropy (TEE)~\cite{Levin2006,Kitaev2006} cannot suitably capture 
the universal character since the entanglement entropy constituting the TEE includes classical and 
quantum correlations~\cite{Werner1989,Winter_2005}. 
Thus, to characterize a mixed state quantum topological order, the TEE may be insufficient.
On the other hand, there are some physical quantities to capture quantum correlation for 
mixed states~\cite{Horodecki_family}. 
One of the efficient candidate is the negativity~\cite{peres1996,Vidal_2002,Plenio_2005}.
Recently for the mixed state topological order, it is suggested that
the combination of the negativity analogous to the TEE, 
named topological entanglement negativity (TEN), can be a good diagnosis to characterize universal 
properties of the mixed state topological order, especially for the imTO~\cite{Fan_2024,Wang_2025, KOH2025_v1}. However, the universal characterization for the imTO and other mixed topological orders by focusing
on the TEN remains an open problem.

\subsection*{Motivation of this work and summary of the main results}
This work studies the color code under decoherence described by operators residing on links and 
examine how the color code changes into non-trivial mixed states without a counter part in pure states. 
The starting point of this study is the fact: 
based on the anyon picture, the color code is connected to two copies of the TC \cite{Bombin_2012NJP,Kubica_2015,Aloshious2019,Kesselring2024}, that is
$$
\text{color code} \simeq \text{toric code} \times \text{toric code}.
$$
This fact gives us an intuitive physical expectation; By some measurements and/or decoherence (noise), 
the pure color code corresponding to the double TC system may be reduced into the single TC. 
The reduced state is a  mixed state with a topological order.
Furthermore, if such a phenomena occurs, how the change occurs is a very interesting issue 
and it is important to verify obtained qualitative observation by the numerical methods.

There is a work motivated by a similar idea to the above \cite{PhysRevResearch.6.L042063}.
There, measurement only circuit in a honeycomb lattice was numerically studied in
the stabilizer formalism with recoding outcomes. 
The states under study stay in pure states, and the way in which a color code state changes to a TC state
is observed through the TEE. 
In fact, the TEE changes from $2\ln 2$ to $\ln 2$ as increasing the strength of projective measurement. 

The above study focused on pure states, for which the TEE is an efficient tool.
Beyond the pure-state framework, in this work, we study the mixed states in the color code 
under a specific type of decoherence. 
Through the practical investigation, we clarify the emergence of the imTO through the anyon picture. 
There, it is important to judge which modular or nonmodular theory the anyon data in the decohered mixed 
states exhibit. 
The modular theory is a theory in which every anyon in the theory has non-trivial braiding with 
at least one other anyon. 
The nonmodular theory is a theory in which at least one anyon has only trivial braiding with
all other anyons, which is referred to as transparent.
The imTO is defined to be in the nonmodular theory~\cite{Sohal2025,Ellison_PRXQuantum_2025}.
In particular, this judgment can be easily obtained by applying gauging out scheme as discussed in \cite{Sohal2025,Ellison_PRXQuantum_2025}.

In this work, we explicitly 
elucidate the nonmodular theory (premodular theory) in the decohered color code. 
Then, by using the notion of anyon proliferation, 
we show that the modular theory obtained by factoring the the transparent-anyon sector coincides with
the anyon content of a single TC.
We conclude that the decohered color code is the imTO including the TC modular sector since there is an explicit transparent anyon related to the emergent strong $1$-form symmetry. 

For the modular theory, we expect that the TEN is a good diagnose for the dcohered color code \cite{Cai2026}. 
Indeed, we expect the change of TEN: $2\ln 2 \longrightarrow \ln 2$. 
This change can be analytically examined by the practical calculation of the negativity for small systems.
Moreover, we numerically investigate the change of the color code under the stochastic types of decoherence 
by using the efficient algorithm of the stabilizer formalism for large 
systems \cite{gottesman1998,aaronson2004} . 
In fact, we numerically observe the TEN and find  a clear change of the value of the TEN, $2\ln 2 \longrightarrow \ln 2$ 
in a intermediate strength of the decoherence and the clear appearance of the imTO characterized by TEN.
To our best knowledge, this is the first study investigating in detail the negativity and TEN by systematic
numerical methods.
In fact, we observe specific and unexpected behavior of them in the intermediate regime between
the color code and TC, details of which will be reported in the later sections.

In this work, beyond the general and/or abstract construction and the limited examples of the topological 
order (stabilizer code) under decoherence \cite{Sohal2025,Wang_2025,Ellison_PRXQuantum_2025}, we clearly 
show the imTO originated from the color code under the decoherence. 
Indeed, we consolidate the notion of the imTO through the study of the color code under a specific type of 
decoherence and show the richness of concrete imTO's.\\

The plan of the paper is as follows. 
In Sec.~II, we introduce the color code, decoherence and the stabilizer formalism. 
Then, we explain the properties of the color code and its anyon contend.
In Sec.~III, we consider the maximal decoherence limit of the color code.
Based on the stabilizer formalism, we discuss how the color code state changes by observation of red-link
$XX$ operators by using the gauging out perspective \cite{ellison2023,Sohal2025}. 
We further study the anyon theory (anyon data) for the decohered state, where we employ the notion of 
he anyon proliferation \cite{Sohal2025,Ellison_PRXQuantum_2025}.
Then, we discuss 1-form symmetries existing in the decohered state. 
In Sec.~IV, we introduce the entanglement negativity and the TEN. 
By using the results in Sec.III, we give the explicit characterization of the decohered state through the TEN 
and show the analytical verification through the practical formula in the stabilizer formalism. 
In sec.~V we move to the numerical study by the efficient stabilizer algorithm \cite{gottesman1998,aaronson2004}, 
where we apply stochastic decoherence to the color code and observe both the TEN and the negativity. 
We find that for intermediate strengths of the decoherence, both the TEN and negativity exhibit 
large but smooth changes, whereas their variances are large and system-size independent. 
We also find that calculations of the negativity support a scaling law, that is, the negativity exhibits 
a area law scaling form with the constant value of the TEN.

Section~VI is devoted to discussion and conclusion.

\begin{figure*}[t]
\begin{center}
\includegraphics[width=18cm]{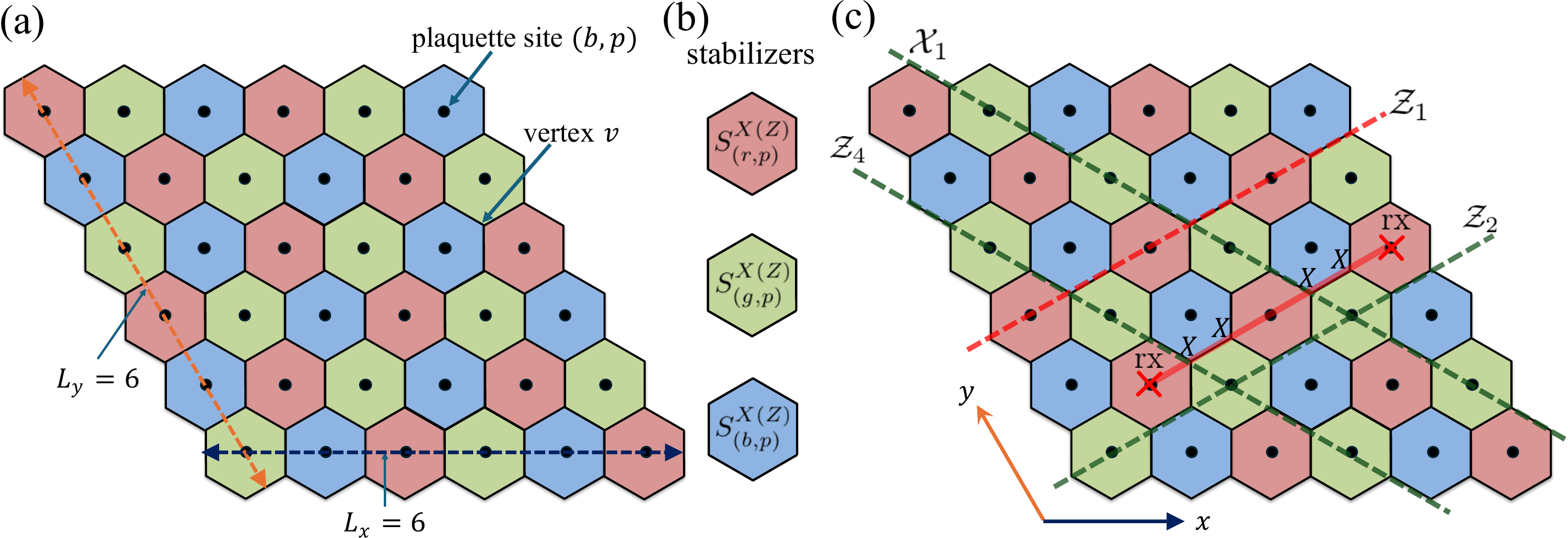}
\end{center}
\caption{Schematic image of color code system. (a) Image of the lattice and the definitions of plaquette site, vertex and length of the system in this honeycomb lattice. 
The qubit resides on the vertex $v$.
When we take this parallelogram-shaped honeycomb lattice and impose periodic boundary conditions (corresponding to torus geometry) by identifying the top with the bottom and the left with the right. The system becomes a honeycomb torus lattice with $L_x = L_y = 6$.
(b) Three colored stabilizers. The plaquette stabilizer are classified into three color; red, green and blue. 
The Pauli operator $X$ or $Z$ are resided on each corners (corresponding to the vertex) of the hexagon. These plaquette stabilizers are building blocks of the stabilizer Hamiltonian $H_{\rm CC}$. (c) Example of the pair of logical operator $\mathcal{X}_1$ (one of dashed green line in y-direction) and $\mathcal{Z}_1$ (the dashed red line). The logical operators crosses at odd-times even through however deformed these non-contractible loop are. Thus, $\mathcal{X}_1$ and $\mathcal{Z}_1$ are anti-commute each other. The red-solid line represents an open string $X$ operator passing through red-links. This is an example for the creation of anyon excitation where two ${\rm rx}$ anyons are created at the ends of the string operator. The other types of logical $Z$ operators  $\mathcal{Z}_2$ (The another of dashed green lines in $y$-direction) and $\mathcal{Z}_4$ (the dashed green line in $x$-direction) are displayed. $\mathcal{X}_1$ and $\mathcal{Z}_4$ are commute even if these lines share same vertices under some deformations since the number of the shared vertices is even.
}
\label{Fig1}
\end{figure*}
\section{Color code and decoherence}
We start to introduce the system, that is, the color code, its stabilizer formalism and 
decoherence that will be studied in this work. 
We first consider a honeycomb lattice on a torus $T_2$ with the two spatial directions $x$ and $y$
(See Fig.\ref{Fig1}(a)). 
Then, spin $1/2$, qubit, is put on each vertex on the lattice. 
Throughout this work, the spatial length unit is the distance between centers 
of the adjacent hexagons.
We also comment on the system size and geometry considered in this work.
As we take a parallelogram-shaped honeycomb lattice as shown in Fig.~\ref{Fig1}(a) and impose periodic 
boundary conditions by identifying the top with the bottom and also the left with the right, the lattice forms a 
torus. 
This honeycomb lattice on the torus with $L_x = L_y = 6$ is displayed in Fig.~\ref{Fig1}(a).

Then, a way of coloring hexagons with three kinds of color, red, green and blue, is introduced as shown in
Fig.~\ref{Fig1}(b), in which adjacent hexagons  have different colors with each other.
We also classify the links of the lattice: In this honeycomb lattice, each link connects two vertices that belong to hexagons of the same color. 
As a result, the links can be classified into three types, as shown in Fig.~\ref{Fig1}. 
We refer to them as red links, green links, and blue links, respectively.

\subsection{Stabilizer Hamiltonian}
In this work, we study the properties of the color code under specific decoherence, which 
connects the color code to the TC \cite{Kitaev_1997}.
Hamiltonian of the color code is given
as \cite{Bombin2006,Bombin2012,Bombin2015,Yoshida_2015,Kubica_2015},
\begin{eqnarray}
H_{\rm CC}=-\sum_{(p,c)}\biggl[S^X_{(c,p)}+S^Z_{(c,p)}\biggr],
\label{CC1}
\end{eqnarray}
with operators 
\begin{eqnarray}
S^X_{(c,p)}\equiv \prod_{v\in (c,p)}X_v,\:\:\:\:\ S^Z_{(c,p)}\equiv \prod_{v\in (c,p)}Z_v,
\label{stab1}
\end{eqnarray}
where,$X_v$, $Z_v$ are Pauli operators residing on the vertex $v$, $c$ refers to a ``color'' namely red($r$), green($g$) 
and blue($b$), taking $c=r,g,b$, and $p$ denotes a plaquette (hexagon). 
In this paper, we use plaquette and hexagon interchangeably.
We use the following labels; 
$(c,p)$ to represent all plaquettes on the honeycomb lattice without redundancy. 
$v$'s are the vertices around the plaquette $(c,p)$. 
The image of these plaquette terms are shown in Fig.~\ref{Fig1}(b).
The Hamiltonian is called ``stabilizer Hamiltonian'' since all terms are commute with each other and they
play a role of {\it stabilizers} and are elements of the Pauli group $\mathcal{P}$ \cite{Wen_text}.
The ground states of the Hamiltonian $H_{\rm CC}$ in Eq.~(\ref{CC1}), $|{\rm GS}\rangle$, are simply given as
\begin{eqnarray}
S^X_{(c,p)}|{\rm GS}\rangle = S^Z_{(c,p)}|{\rm GS}\rangle = |{\rm GS}\rangle,
\label{GS1}
\end{eqnarray}
for all $(c,p)$, and equations in (\ref{GS1}) are conditions that $|{\rm GB}\rangle$ is the color-code state or simply
color code expressed in the stabilizer formalism.
The quantum code and properties of anyons encoded within $H_{\rm CC}$ in Eq.~(\ref{CC1})
are reviewed in detail in \cite{Kesselring2024}, which will be explained later on. 

On the torus geometry, there are redundancy among the stabilizers, i.e., 
the terms of the Hamiltonian $H_{\rm CC}$, such as \cite{Bombin2006}:
\begin{eqnarray}
&&\prod_{v\in (c=r,p)}S^\alpha_{(c,p)}=\prod_{v\in (c=g,p)}S^\alpha_{(c,p)}=\prod_{v\in (c=b,p)}S^\alpha_{(c,p)}\nonumber\\
&&=\prod_{v:\mbox{all}}\alpha_v,
\label{const1}
\end{eqnarray}
where $\alpha=X$ and $Z$. 
From the above relationship, there are four identities:
\begin{eqnarray}
\biggr[\prod_{v\in (c=r,p)}S^\alpha_{(c,p)}\biggl]\cdot \biggr[\prod_{v\in (c=g,p)}S^\alpha_{(c,p)}\biggl] = I,\label{const2}\\
\biggr[\prod_{v\in (c=r,p)}S^\alpha_{(c,p)}\biggl]\cdot \biggr[\prod_{v\in (c=b,p)}S^\alpha_{(c,p)}\biggl] = I.
\label{const3}
\end{eqnarray}
Then, the number of the independent stabilizers called stabilizer generators is $N_v-4$, where $N_v$ is the total 
number of the vertex. 
Thus, this redundancy of stabilizer generators gives the ground state degeneracy of the Hamiltonian $H_{\rm CC}$ 
such as  $2^{N_v-(N_v-4)}=2^4=16$. 
This means that the ground state can encode four logical (quantum) qubits~\cite{Bombin2006}.
There exist eight non-contractible logical operators from the viewpoint of homology class, which can 
manipulate these logical four qubits. 
Here, we set the labels $x$ and $y$ as two fundamental cycles of the torus.
Then, non-contractible loop operators are given by
\begin{eqnarray}
\mathcal{Z}_1&=&\prod_{v\in n\ell(r,x)}Z_{v},\:\:
\mathcal{Z}_2=\prod_{v\in n\ell(g,x)}Z_v,\nonumber\\
\mathcal{Z}_3&=&\prod_{v\in n\ell(r,y)}Z_{v},\:\:
\mathcal{Z}_4=\prod_{v\in n\ell(g,y)}Z_v,\nonumber\\
\mathcal{X}_1&=&\prod_{v\in n\ell(g,y)}X_{v},\:\:
\mathcal{X}_2=\prod_{v\in n\ell(r,y)}X_v,\nonumber\\
\mathcal{X}_3&=&\prod_{v\in n\ell(g,x)}X_{v},\:\:
\mathcal{X}_4=\prod_{v\in n\ell(r,x)}X_v,
\label{logical_OPs}
\end{eqnarray}
where the label $n\ell(c,x(y))$ represents a non-contractible loop-pass on the torus, tracing on the $c$-color links and in the $x(y)$-direction. 
A schematic image of the logical operators is shown in Fig.~\ref{Fig1}(c).
These non-contractible operators constitute four logical operator algebra:
\begin{eqnarray}
\{\mathcal{Z}_{\beta},\mathcal{X}_{\beta}\}=0, \:\:\mbox{with}\:\:\beta=1,2,3,4,
\label{4log_algebra}
\end{eqnarray} 
where $\beta$ corresponds to the label of encoded logical qubits. 
These anti-commutative relations of the logical operator pairs preserve that the code is quantum.

In addition, we comment on the symmetry aspect. The logical operators $\mathcal{X}_{\beta}$ and $\mathcal{Z}_{\beta}$ can be regarded as 
the non-trivial action of $Z_2$ 1-form symmetry \cite{Zohar2009,NUSSINOV_2009,Gaiotto_2015,mcgreevy2023}, 
such as $\mathcal{Z}_{\beta}H_{\rm CC}(\mathcal{Z}_{\beta})^\dagger=H_{\rm CC}$. 
These symmetries cannot be satisfied simultaneously in the states of the color code; that is, 
the states cannot be a simultaneous eigenstate of all these symmetry operators. 
In other words, all of these symmetries cannot be imposed at the same time, leading to a situation 
analogous to ’t Hooft anomaly\cite{thooft1978}, called ``mixed-anomaly''\cite{Sohal2025}. 
The four anti-commutative pairs can be regarded as the presence of four mixed-anomaly in the theory. 
From this perspective, the resulting ground-state degeneracy of $H_{\rm CC}$ can also be understood as 
a consequence of this anomaly.

\subsection{Decoherence}
This work focuses on two-body operators of decoherence, described by the following quantum channel \cite{Nielsen2011}
\begin{eqnarray}
&&\mathcal{E}^{XX}=\prod_{(v_r,v'_r)}\mathcal{E}^{XX}_{(v_r,v'_r)},\nonumber\\
&&\mathcal{E}^{XX}_{(v_r,v'_r)}[\rho]=(1-p_r)\rho+p_r (X_{v_r}X_{v'_r})\rho (X_{v_r}X_{v'_r}),\nonumber\\
\label{deco_XX}
\end{eqnarray}
where $(v_r,v'_r)$ represents edge-vertices of the red-link. 
The decoherence is applied solely to all red links stochastically and  $p_r$ is the strength of decoherence taking 
$0\leq p_r\leq 1/2$. 
This channel of the decoherence makes a pure state mixed for any finite value of  $p_{r}$. 
In this work, we prepare the pure ground states of the color code Hamiltonian $H_{\rm CC}$ denoted as 
$\rho_{\rm CC}\equiv |{\rm GS}\rangle\langle {\rm GS}|$
(the detail description of the stabilizer generators is shown in later) and study the entanglement 
properties of the mixed state $\mathcal{E}^{XX}[\rho_{\rm CC}]$.

In particular for the $p_r=1/2$ limit called maximal decoherence or fixed point, the decoherence 
corresponds to the projective measurement of $X_{v_r}X_{v'_r}$ {\it without recoding outcomes}. 
We note that this process is different from the projective measurement carried out in a similar 
system~\cite{PhysRevResearch.6.L042063}.
Indeed, the state $\mathcal{E}^{XX}[\rho_{\rm CC}]$ is a  mixed state even for $p_r=1/2$. 

This work mainly proceeds based on the stabilizer formalism, and
in the analytical analysis, we mostly focus on the fix point case $p_r=1/2$. 
By means of the Gottesman-Aaronson efficient stabilizer algorithm \cite{aaronson2004,weinstein2022},
we efficiently carry out numerics to observe how the color code changes to the TC under
the decoherence.

\subsection{Stabilizer formalism}
The color code system with or without decoherence can be efficiently investigated by the stabilizer 
formalism~\cite{gottesman1998,Nielsen2011}. 
In particular, the stabilizer representation of the density matrix gives us the information on decohered states 
directly including the behavior and properties of anyons in mixed states. 
In addition, the stabilizer formalism also serves as the formalism describing the gauging out procedure 
for decoherence process as we explain it later on~\cite{ellison2023,Sohal2025}.

We begin with describing the color code. 
It is represented by a set of stabilizer generators extracted from the Pauli group denoted by $\mathcal{S}_{\rm CC}$. 
As one of the simplest representations for it, we chose $\mathcal{S}_{\rm CC}$ given by
\begin{eqnarray}
\mathcal{S}_{\rm CC}&=&\{S^X_{(r,p)}\}'+\{S^X_{(g,p)}\}'+\{S^X_{(b,p)}\}\nonumber\\
&+&\{S^Z_{(r,p)}\}'+\{S^Z_{(g,p)}\}'+\{S^Z_{(b,p)}\}
\equiv \{g_{\ell}\}.
\end{eqnarray}
Here, it should be noted that all plaquette stabilizers shown in Fig.~\ref{Fig1}(b) are not linearly-independent. There are four identities between them as shown in Eqs.~(\ref{const2}) and (\ref{const3}). 
Then, to fix an appropriate set $\mathcal{S}_{\rm CC}$, it is necessary to remove two arbitrary plaquette operators from 
the $X$-type and two from $Z$-type plaquette stabilizers, respectively. 
In the practical calculation and consideration, we remove one plaquette stabilizer from each of four subsets, 
i.e., $S^X_{(r,p)}$, $S^X_{(g,p)}$ and $S^Z_{(r,p)}$, $S^Z_{(g,p)}$ are removed. 
These manipulation is indicated by the representation of subset $\{\cdot\}'$. 
For notational convenience, these generators are re-named as $g_{\ell}$ (omitting the color and the type of 
Pauli operators), where $\ell=0,1,\cdots, N_v-5$.

Based on this  choice of $\mathcal{S}_{\rm CC}$, the $2^4$ degenerate states of color code can be represented
by the density matrix $\rho_{CC}$ as the product of the projectors given by
\begin{eqnarray}
\rho_{\rm CC}\equiv \frac{1}{2^{N_v-4}}\prod^{N_v-5}_{\ell_=0} \biggr[\frac{I+g_{\ell}}{2}\biggr],
\end{eqnarray}
where $I$ is the identity operator.
It is obvious that $\rho_{\rm CC}$ satisfies the stabilizer condition 
$g_\ell\rho_{\rm CC}=\rho_{\rm CC}g_\ell=\rho_{\rm CC}$
for all $\{g_\ell\}$.

 In Appendix A, we explain how the local decoherence $\mathcal{E}^{XX}_{(v_r,v'_r)}$ for $p_r=1/2$ practically works on
$\rho_{\rm CC}$ in the stabilizer formalism. 
This manipulation is implemented in the numerical  algorithm.
Practically, this local operation deletes two stabilizers, $S^Z_{(r,p)}$ and $S^Z_{(r,p')}$,
where $p$ and $p'$ are red plaquettes sharing the vertices $v$ and $v'$, respectively.

\subsection{Anyons in color code}
In this subsection, we summarize content of the anyon in the color code. 
The code has $16$ type anyonic excitations; $9$ non-trivial bosonic anyons and $6$ non-trivial fermionic anyons 
and the remaining one is the trivial one, the vacuum \cite{Kargarian_2010,Kubica_2015,Kesselring2024,Haghighi2025}. 

First, we explain the $9$ non-trivial bosonic anyons. 
These are classified by two indices; i.e., color and Pauli operator $X$, $Y$ and $Z$, and the anyons
are labeled as  $({\rm rX},{\rm rY},{\rm rZ})$, $({\rm gX},{\rm gY},{\rm gZ})$ and $({\rm bX},{\rm bY},{\rm bZ})$.
These anyons also have fusion rules the detail of which is explained in \cite{Kesselring2024}.
As an example of the excitation process, see Fig.~\ref{Fig1}(c), the red-solid $X$ Pauli open string operator
on the red-links creates a pair of ${\rm rX}$ anyon residing on the red plaquettes 
where the open $X$-string ends and the stabilizer condition $S^Z_{(r,p)}=1$ is broken. 
In other words, the string excites a pair of red color plaquettes.
The other types of the non-trivial anyons can be created in a pair in the same way by using different Pauli 
operator and setting on the open string on color-fixed links. 
In addition, as another example of the excitation process, we apply the local $X$ Pauli operation to a single
vertex $v$ that creates ${\rm rX}$, ${\rm gX}$ and ${\rm bX}$ in three surrounding plaquettes with different colors representing
the fusion process~\cite{Kesselring2024}. 
Another Pauli component anyons can be created in the same way.

Next, we briefly mention the $6$ non-trivial fermionic anyons. 
We denote them by ${\rm f}_\alpha$ with $\alpha=1,2,\cdots,6$, and  ${\rm f}_\alpha$'s are created by fusing two non-trivial 
bosonic anyons~\cite{Kargarian_2010,Kesselring2024}:
\begin{eqnarray}
{\rm f}_1 &=& {\rm rX} \times {\rm bZ} = {\rm rY} \times {\rm gZ} = {\rm gX} \times {\rm bY},\nonumber\\
{\rm f}_2 &=& {\rm rZ} \times {\rm bX} = {\rm rY} \times {\rm gX} = {\rm gZ} \times {\rm bY},\nonumber\\
{\rm f}_3 &=& {\rm bZ} \times {\rm gY} = {\rm gX} \times {\rm rZ} = {\rm bX} \times {\rm rY}, \nonumber\\
{\rm f}_4 &=& {\rm rZ} \times {\rm gY} = {\rm gX} \times {\rm bZ} = {\rm rX} \times {\rm bY}, \nonumber\\
{\rm f}_5 &=& {\rm rX} \times {\rm gY} = {\rm bX} \times {\rm gZ} = {\rm bY} \times {\rm rZ}, \nonumber\\
{\rm f}_6 &=& {\rm bX} \times {\rm gY} = {\rm rX} \times {\rm gZ} = {\rm rY} \times {\rm bZ}, 
\end{eqnarray}
where $\times$ denotes fusion. 
Here, even (odd)-number fermionic anyon trivially braids with odd (even)-number fermionic anyon, and 
even (odd)-number fermionic anyon non-trivially braids with even (odd)-number fermionic anyon. 
The non-trivial braiding factor is only $(-1)$ since we are considering an Abelian anyon theory.

The above set of $16$ anyons, $\mathcal{A}_{\rm CC}$, equals two  sets of 
TC anyon, $\mathcal{A}_{TC}$, that is, $\mathcal{A}_{\rm CC}\cong\mathcal{A}_{TC}\times \mathcal{A}_{TC}$, where $\mathcal{A}_{TC}= \{{\rm 1}, {\rm e}, {\rm m}, {\rm f}={\rm e}\times {\rm m}\}$~\cite{kitaev2003}. 
In this sense, the color code can be regarded as two decoupled TC's \cite{Bombin_2012NJP,Kubica_2015,Kesselring2024}. 
There are many isomorphism how the color code anyons are related to the anyons of the double decoupled TC. 
One of the examples is herein shown~\cite{Kesselring2024}. 
First, the bosonic anyons of the color code are related to composite of anyons of the TC's as
\begin{eqnarray}
{\rm rX} &=& ({\rm e},{\rm 1}),\;\; {\rm rY} = ({\rm e},{\rm m}),\;\; {\rm rZ} = ({\rm 1},{\rm m}),\nonumber\\
{\rm gX} &=& ({\rm e},{\rm e}),\;\; {\rm gY} = ({\rm f},{\rm f}),\;\; {\rm gZ} = ({\rm m},{\rm m}),\nonumber\\
{\rm bX} &=& ({\rm 1},{\rm e}),\;\; {\rm bY} = ({\rm m},{\rm e}),\;\; {\rm bZ} = ({\rm m},{\rm 1}).
\label{color_TC2_anyon}
\end{eqnarray}
Second, the fermionic anyons of the color code are also related to anyons of the TC's as
\begin{eqnarray}
{\rm f}_1 &=& ({\rm f},{\rm 1}),\;\; {\rm f}_2 = ({\rm e},{\rm f}),\;\; {\rm f}_3 = ({\rm m},{\rm f}),\nonumber\\
{\rm f}_4 &=& ({\rm 1},{\rm f}),\;\; {\rm f}_5 = ({\rm f},{\rm e}),\;\; {\rm f}_6 = ({\rm f},{\rm m}).
\end{eqnarray}
Note that a single color code anyon is not generated as a composite object of two independent TC anyons. 
Rather, these relations should be understood purely as a mapping between the two descriptions.

\section{Decoherence effect and gauging out}
In this section, we study the effects of the decoherence $\mathcal{E}^{XX}$ with $p_r=1/2$ in Eq.~(\ref{deco_XX}) 
on the color code.
From the view point of anyons, the decoherence $\mathcal{E}^{XX}$ induces the proliferation of the ${\rm rX}$ anyon \cite{Wang_2025,Sohal2025}, the notion of which is different from anyon condensation in pure states 
since the proliferation anyon cannot be regarded as a condensed object that is ``absorbed''
by the vacuum~\cite{Sohal2025}.

\subsection{Logical operator perspective} 
We investigate how the color code varies under the decoherence. 
We give a qualitative but efficient observation how the decoherence $\mathcal{E}^{XX}$ acts on the eight logical 
operators in Eq.~(\ref{logical_OPs})~\cite{Sohal2025}.
We apply $\mathcal{E}^{XX}$ to the $Z$-logical operators to easily find 
\begin{eqnarray}
\mathcal{E}^{XX}[\mathcal{Z}_{2}]=0,\:\: \mathcal{E}^{XX}[\mathcal{Z}_{4}]=0,
\end{eqnarray} 
as ${\cal Z}_2$ and ${\cal Z}_4$ anti-commute with some of $XX$ operators on red link 
since a single red link shares only a single vertex with the loop operators of ${\cal Z}_2$ and ${\cal Z}_4$ as shown 
in Fig.~\ref{Fig1} (c).
The others are invariant under the operation of  $\mathcal{E}^{XX}$. 
Therefore, the logical operators $\mathcal{X}_{2}$ and $\mathcal{X}_{4}$ lose anti-commutative partner, 
while the other two keep the anti-commutative logical operators,
\begin{eqnarray}
\{\mathcal{Z}_{1},\mathcal{X}_{1}\}=0,\:\:
 \{\mathcal{Z}_{3},\mathcal{X}_{3}\}=0.
\label{2log_algebra}
\end{eqnarray} 
From the perspective of the encoded qubits, this observation implies that the logical qubits originally present 
in the color code are affected by the decoherence $\mathcal{E}^{XX}$ in such a way that two of them become classical bits, 
while the remaining two continue to exist as encoded quantum qubits.

Intuitively, it is expected that the remaining logical qubits correspond to the two logical qubits 
of the TC defined on a torus. 
In this picture, among the two TC layers corresponding to the original color code, one loses its quantum 
nature due to the proliferation of rX-anyon and effectively reduces to two classical bits
(a similar situation often occurs~\cite{Wang_2025,Sohal2025}), so that one TC disappears, 
while the other survives and preserves the two logical qubits.

More precisely from the viewpoint of the anyon theory, the fact that $\mathcal{X}_{2}$ and 
$\mathcal{X}_{4}$ do not accompany anti-commutative pair operators indicates the existence of the transparent anyons, 
which means that the decohered state is categorized in non-modular theory. 
That is, the decohered state can be the imTO \cite{Sohal2025}.

The above observations suggest that the decoherence $\mathcal{E}^{XX}$ gives rise to a mixed-state TC possessing two 
classical bits and two quantum bits, implying a form of mixed topological order. 
An efficient quantity to characterize this mixed topological order stemming from the partially remaining 
TC is the TEN, which will be introduced later.

\subsection{Gauging out}
Based on the stabilizer formalism, the decoherence to the color code can be understood as a
gauging out process~\cite{ellison2023,Sohal2025}. 
The gauging out changes the original the stabilizer group of the color code to the new one corresponding
to the decohered mixed state. 
The set of the stabilizers obtained by the gauging out gives various information about the emergent
decohered mixed state under $\mathcal{E}^{XX}$.
 
Following the subsystem code formalism utilizing the gauge group \cite{Sohal2025}, we set the gauge 
group of the target system by combining the stabilizer group $\mathcal{S}_{\rm CC}$ with applied $XX$ operators 
on the red links as follows~\cite{Sohal2025},
\begin{eqnarray}
\mathcal{G}_{\rm CC}&=&\{i\}+\{ X_{v_r}X_{v'_r}\}+\{S^X_{(r,p)}\}'\nonumber\\
&+&\{S^Z_{(r,p)}\}'+\{S^Z_{(g,p)}\}'+\{S^Z_{(g,p)}\}.
\label{Gauge_group1}
\end{eqnarray}
Please note that the gauge group $\mathcal{G}_{\rm CC}$ is non-Abelian.
The set of the stabilizers of the decohered state $\rho_{D}$ is given by the center of $\mathcal{G}_{\rm CC}$, $\mathcal{Z}(\mathcal{G}_{\rm CC})\equiv \mathcal{S}_D$ such as,
$\rho_D\propto \prod_{\ell}\frac{I+g^{D}_\ell}{2}$ where $\{g^D_{\ell}\}$ are a set of  generators of $\mathcal{S}_D$.

By following \cite{ellison2023}, we can find the center as follows {\it for the maximal decoherence},
\begin{eqnarray}
\mathcal{S}_D=\mathcal{Z}(\mathcal{G}_{\rm CC})&=&\{i\}+\{\mathcal{X}_2,\mathcal{X}_4\}\nonumber\\
&+&\{S^X_{(r,p)}\}'+\{S^X_{(g,p)}\}'+\{S^X_{(b,p)}\}\nonumber\\
&+&\{S^Z_{(g,p)}\}'+\{S^Z_{(b,p)}\}.
\label{ZCC}
\end{eqnarray}
Note that the set of stabilizer generator $\mathcal{S}_D$ includes the non-contractible generators corresponding to 
the logical operators $\mathcal{X}_2$ and $\mathcal{X}_4$, which lost the anti-commutative partners. 
Here, we follow the convention that includes the non-contractible operators  $\mathcal{X}_2$ and $\mathcal{X}_4$ in the 
stabilizer group as in Ref.~\cite{ellison2023}. 
Another comment is in order: Note that the choice of generators in Eq.~(\ref{ZCC}) is not unique.
One can equivalently choose another generators set related by products of stabilizers.
In particular, suitable products of red $Z$-plaquette operators can also be taken as generators, 
which correspond to extended red-$Z$ cluster operators appearing in the stabilizer update picture, 
which are used later on.

The set of the decohered stabilizer generators $\mathcal{S}_D$ contains nonlocal generators 
$\mathcal{X}_2$ and $\mathcal{X}_4$ in its center. 
As stated earlier, together with the disappearance of their anti-commuting partners $\mathcal{Z}_2$ and $\mathcal{Z}_4$, 
this indicates the emergence of a transparent anyon in the decohered state. 
Since the present decoherence proliferates the ${\rm rX}$ anyon, this transparent anyon is the ${\rm r}X$-anyon.

On the other hand, it is not so easy to directly identify the TC-like sector in the decohered state solely 
from Eq.~(\ref{ZCC}).  
In other words, careful look at the structure of the stabilizer group is necessary to identify
the surviving TC.
Nevertheless, the existence of the transparent sector already implies that the resulting anyon theory is
non-modular. 
In this sense, Eq.~(\ref{ZCC}) captures an essential signature of the non-modular (premodular, more precisely) 
nature of the decohered state. 
Furthermore, the surviving nonlocal operators $\mathcal{X}_2$ and $\mathcal{X}_4$ can be viewed as reflecting a residual classical 
loop sector associated with this transparent excitation. 
As will be discussed later, the extraction of TC-like topological signatures can be carried out 
by observing the TEN, which provides strong supporting evidence for TC-like topological signatures.

\subsection{Identification of  toric-code type anyon structure in decohered state}
In the previous subsection, we saw that the ${\rm rX}$ anyon proliferates in the decohered state and obtain 
the set of the stabilizer generator $\mathcal{S}_D$. 
Here, we discuss the properties of anyons in decohered states by following the discussions 
in \cite{Kesselring2024,Sohal2025}.

In the initial color code, total 16 anyons exist (including the vacuum anyon). 
Then, we consider the decoherence $\mathcal{E}^{XX}$ inducing the proliferation of ${\rm rX}$ anyon. 
The remaining 14 (non-trivial) anyons in the color code are two types; confined or deconfined. 
If an anyon non-trivially braids to the anyon ${\rm rX}$, it is confined, and if trivially braids, it is deconfined. 
In the decohered states, the deconfined anyons plays a key role characterizing the state. 
Here, in the decohered state with proliferation of the ${\rm rX}$ anyon, 6 non-trivial deconfined anyons appear, 
the set of which is 
\begin{eqnarray}
\mathcal{A}^{\rm rX}_{dec}&=&\{{\rm gX}=(e,e),{\rm bX}=(1,e),
{\rm rY}=(e,m),\nonumber\\
&&{\rm rZ}=(1,m),
{\rm f}_2=(1,f),
{\rm f}_3=(e,f)\}. 
\end{eqnarray}
Here, we introduce an equivalence among them by employing the transparent anyon ${\rm rX}$. 
That is, we set the following equivalence
\begin{eqnarray}
{\rm gX}&=&{\rm bX}\times {\rm rX}\longrightarrow {\rm gX}\sim{\rm bX},\nonumber\\
{\rm rY}&=&{\rm rZ}\times {\rm rX}\longrightarrow {\rm rY}\sim{\rm rZ},\nonumber\\
{\rm f}_2&=&{\rm f}_3\times {\rm rX}\longrightarrow {\rm f}_2\sim{\rm f}_3. 
\end{eqnarray}
This is the same identification in the case of the anyon condensation caused by 
measurement instead of decoherence~\cite{Kesselring2024}. 
Also, we consider the group $\{{\rm rX}\}\cup \mathcal{A}^{\rm rX}_{dec}\equiv \mathcal{A}_{D}$.
Then, $\{{\rm rX}\}$ is normal subgroup in $\mathcal{A}_{D}$ since our theory is Abelian. 
Thus, we can consider a quotient group given by $\mathcal{A}_{D}/\{{\rm rX}\}$.
Here, we identify the quotient group as \cite{Ellison_PRXQuantum_2025},
\begin{eqnarray}
\mathcal{A}_{D}/\{{\rm rX}\} \cong \{\mbox{toric-code-type anyon structure}\}.
\label{TC_modular_part}
\end{eqnarray}
Note that as the modular quotient of the deconfined sector to the transparent anyon $\{{\rm rX}\}$, a toric-code-type anyon structure emerges. This modular part by factoring the transparent anyon ${\rm rX}$ gives us the expectation that the decohered state by $\mathcal{E}^{XX}$ possesses the toric-code-like sector in the decohered mixed state.

In addition, we give careful comments: 
\begin{enumerate}
\item Although 
${\rm rX}$ is not literally absorbed into the vacuum sector in the gauge-out description \cite{Sohal2025}, 
it acts as a transparent anyon that is invisible to all braiding processes involving the surviving 
deconfined excitations. 
This allows us to quotient the deconfined anyon set by fusion with ${\rm rX}$, leading to an emergent toric-code-type modular structure. In this sense, decoherence induces a nontrivial reorganization of the anyon content, effectively reducing a non-modular theory to a modular one. 
This modular structure $\mathcal{A}_{D}/\{{\rm rX}\}$ characterizes the decohered theory. 
This can give the exact value of the TEN since the modular part $\mathcal{A}_{D}/\{{\rm rX}\}$ directly gives quantum dimension \cite{Cai2026}, shown in later section.
\item
In the present system, the ${\rm rX}$ anyon behaves in a manner analogous to anyon condensation~\cite{Kesselring2024,Ellison_PRXQuantum_2025}. 
However, it is not absorbed into the vacuum sector and instead remains as a transparent anyon. Consequently, the resulting structure is more appropriately understood not as a true anyon condensation, but as a modular quotient of a non-modular theory arising from the presence of the transparent sector.
Physically, this corresponds to the fact that the emergent mixed state contains a set of pure states distinct  only in the number of the ${\rm rX}$ anyons {\it with equal weight}.
\end{enumerate}

\subsection{Aspects of $1$-form symmetry}
Before closing this section, we briefly discuss the $1$-form symmetries of 
the color code~\cite{Gaiotto_2015,mcgreevy2023},  operators of which are defined on a  loop in 2D system. 
For mixed states, there are two types of $1$-form symmetries, namely strong and weak symmetries~\cite{Buca_2012,Albert2014,groot2022}. 
Their definitions and properties are briefly explained in Appendix B. 
In the present case, we can introduce two types of loop operators as the generator of the $1$-form symmetry; 
contractible and non-contractible. 
The non-contractible ones are defined on two different cycles of the torus.\\

\noindent{\underline{Contractible loop case:}} This case has been discussed in detail in 
Ref.~\cite{Sohal2025}. 
We show that the color code has $(\mathbb{Z}_2)^4$ $1$-form strong symmetry. 
Choice of the four contractible loop operators as the generators of 1-form symmetry is not unique. 
Here, we show one interesting choice elucidating how they change from the strong to weak 1-form 
symmetries under decoherence. 
This choice comes from two point: 
(i) the nine bosonic anyons previously shown have fusion rules such as ${\rm rX}\times {\rm gX}={\rm bX}$, ${\rm rZ}\times {\rm gZ}={\rm bZ}$, etc  \cite{Kesselring2024}, 
(ii) There are six kinds of the plaquette operators $S^{X(Z)}_{(c,p)}$ in $\mathcal{S}_{\rm CC}$ and there exist the identities Eq.~(\ref{const1}) 
among them. 
From these observations, the four loop operators are to be introduced as
\begin{eqnarray}
W^{\rm rX}(\gamma_c)=\prod_{(c=g,b,p)\in \Sigma(\gamma_c)}S^{X}_{(c,p)},\nonumber\\
W^{\rm gX}(\gamma_c)=\prod_{(c=g,b,p)\in \Sigma(\gamma_c)}S^{X}_{(c,p)},\nonumber\\
W^{\rm gZ}(\gamma_c)=\prod_{(c=r,b,p)\in \Sigma(\gamma_c)}S^{Z}_{(c,p)},\nonumber\\
W^{\rm bZ}(\gamma_c)=\prod_{(c=r,g,p)\in \Sigma(\gamma_c)}S^{Z}_{(c,p)},
\end{eqnarray}
where $\gamma_c$ is a contractible loop on the torus and $\Sigma(\gamma_c)$ represents the set of plaquettes within $\gamma_c=\partial \Sigma$.
For the set of the stabilizer generator $\mathcal{S}_{\rm CC}$, we directly observe that each of the above four operators 
represents a 1-form strong symmetry such as $W^{\rm rX}(\gamma_c)\rho_{\rm CC}=\rho_{\rm CC}W^{\rm rX}(\gamma_c)=\rho_{\rm CC}$.

On the other hand, we observe how the symmetry changes from the color code $\rho_{\rm CC}$ to
$\mathcal{E}^{XX}(\rho_{\rm CC})$ described by the set $\mathcal{S}_D$. 
We here see that two 1-from symmetry generated by $W^{\rm gZ}(\gamma_c)$ and $W^{\rm bZ} (\gamma_c)$ become weak
as the plaquette operators $S^Z_{(r,p)}$ disappear by the decoherence, whereas
\begin{eqnarray}
W^{\rm gZ}(\gamma_c)\mathcal{E}^{XX}(\rho_{\rm CC})W^{{\rm gZ}\dagger}(\gamma_c)=\mathcal{E}^{XX}(\rho_{\rm CC}),\nonumber\\
W^{\rm bZ}(\gamma_c)\mathcal{E}^{XX}(\rho_{\rm CC})W^{{\rm bZ}\dagger}(\gamma_c)=\mathcal{E}^{XX}(\rho_{\rm CC}),
\end{eqnarray}
while the strong symmetry condition is not satisfied.
This strong-to-weak ``transition" in the 1-form symmetry can be understood by the unfolding 
picture~\cite{Kubica_2015,Kesselring2024} of the anyon in the color code shown in
Eq.~(\ref{color_TC2_anyon}). 
The decoherence $\mathcal{E}^{XX}$ proliferates ${\rm rX}=(e,1)$ anyon. 
Then, anyons ${\rm gZ}=(m,m)$ and ${\rm bZ}=(m,1)$ have non-trivial braiding with the anyon ${\rm rX}=(e,1)$,
and therefore, they are ``confined''. 
This confined picture is related to the change from the strong 1-form symmetry to weak one. This symmetry-veiwpoint supports the discussion in \cite{Sohal2025}. 
In addition, note the non-trivial braiding is closed in the side of the first TC. 
This fact implies that the properties of anyons in the sector of the second TC does not change, that is, the decohered stat $\mathcal{S}_{\rm CC}$ sustains the properties of the single TC.\\

\noindent\underline{Non-contractible loop case:} 
In addition to the contractible loop, we  briefly comment on the properties 
of the $1$-form symmetry for the non-contractible loops on the torus. 
The generators of the $1$-form symmetry can be regarded as the logical operators $\mathcal{X}_{\alpha}$ and $\mathcal{Z}_{\alpha}$ with
$\alpha=1,2,3,4$.
In the color code, these logical operators constitute four anti-commutative algebra, which can be regarded as mixed anomalies and also  the SSB of the 1-form symmetries. 
This symmetry breaking is the origin of the degeneracy in the code space of the color code.

We next turn to the decohered case $\mathcal{S}_D$. 
The SSB of the $1$-form-symmetry is partially restored and then the 1-form strong symmetries 
with both $\mathcal{X}_2$ and $\mathcal{X}_4$ emerge since the set $\mathcal{S}_D$ includes these generators, that is, $\mathcal{X}_{2(4)}\rho_D=\rho_D\mathcal{X}_{2(4)}=\rho_D$. The original-pair logical operators $\mathcal{Z}_{2(4)}$ are no longer weak nor strong 1-form symmetry 
since $\mathcal{Z}_{2(4)}$ are anti-commute with the generators $\mathcal{X}_{2(4)}$ in $\mathcal{S}_D$. 
On the other hand, the two anti-commutative algebra $\{\mathcal{X}_{1(3)},\mathcal{Z}_{1(3)}\}=0$ (mixed anomalies) remains 
in the decohered state as shown in Sec.III A. 
Thus, these facts tell us that the decohered states $\rho_D$ has ``emergent'' 1-form strong symmetries by $\mathcal{X}_{2(4)}$
and no 1-form weak symmetries for non-contractible loops on the torus.

\section{Entanglement negativity and topological entanglement negativity}
For the mixed states, it is generally difficult to extract genuine pure quantum 
entanglement~\cite{Horodecki_family}. 
However, there is an efficient quantity to capture quantum entanglement in the mixed state, called
negativity~\cite{peres1996,Plenio_2005}. 
In this study, we examine this quantity for the target many-body system. 
We introduce the formulation of the negativity and also the TEN. 
In this section by using the stabilizer formalism \cite{sang2021}, we give some concrete analytical estimation of 
the TEN giving us the valuable information of the topological order in the decohered system.

\subsection{Many-body negativity and its calculation in the stabilizer formalism}
The negativity for a subsystem $A$ in the 2D color code system is defined as 
\begin{eqnarray}
\mathcal{N}_A\equiv \log_2|\rho^{\Gamma_A}|_1=\ln|\rho^{\Gamma_A}|_1/\ln 2,
\end{eqnarray}
where $|\cdot|_1$ represents the trace norm, and $\Gamma_A$ a partial transpose for the basis in the subsystem $A$. 
This quantity is a many-body version of the one examined for a small system  \cite{peres1996,Horodecki_family,Plenio_2005}. 
The negativity is a measure that can identify certain transitions among mixed states and their 
criticality~\cite{lu2020,sang2021,weinstein2022,Anzai_2026}.

Through the stabilizer formalism, the quantity $\mathcal{N}_A$ can be efficiently calculated  \cite{sang2021,shi2021,sharma2022,KOI2023_1}. 
The formula and its detailed derivation were firstly proposed in Refs.~\cite{sang2021,shi2021}. 

As shown in Refs.~\cite{sang2021,shi2021,sharma2022,KOI2023_1}, for finite-size systems, 
$\mathcal{N}_A$ has an explicit formula making use of the stabilizer generators $\{g_{\ell}\}$ for the mixed states, 
which is given by 
\begin{eqnarray}
\mathcal{N}_A=\frac{1}{2}\mathrm{rank}_{\mathbb{F}_2}\tilde{K}_A,
\label{N_A_formula}
\end{eqnarray}
where $\tilde{K}_A$ is a $m \times m$ matrix, $m$ is the total number of the stabilizer generators of a state $\rho$,
$m \le N_v$. 
The factor $1/2$ in Eq.~(\ref{N_A_formula}) arises from counting the Bell-type fundamental quantum correlations 
based on the number of anti-commuting pairs in the stabilizer group~\cite{sang2021}.
Explicitly, the matrix $\tilde{K}_A$ is obtained by searching the anti-commutation pairs for truncated $m$ stabilizer 
generators, where the Pauli-operator components of the generator within the subsystem $A$ remains. 
We denote them by $g^A_{\ell}$ ($\ell=0,1,\cdots, m-1$). 
The truncated stabilizer generators $\{ g^A_{\ell}\}$ are obtained by the two steps: 
First, we represent the stabilizer generators as the binary representation for the basis $X_v$ and $Z_v$ \cite{Nielsen2011} where we ignore the sigh of the stabilizer since the quantities of quantum correlations such as 
the entanglement entropy \cite{Fattal2004} and negativity $\mathcal{N}_A$ do not depend on this sign \cite{Lavasani_NatPhys_2021,Lavasani2021,OKI2024_TC_MoC}. 
Second, we truncate the binary representation vectors of each stabilizer generator to remove the element of 
sites not included in the subsystem A such as 
\begin{eqnarray}
g_{\ell} \longrightarrow g^A_{\ell}=(g^{\ell,X}_0,\cdots, g^{\ell,X}_{k-1} | g^{\ell,Z}_0,\cdots, g^{\ell,Z}_{k-1}),
\end{eqnarray} 
where in $g^{\ell,X(Z)}_{i}$, $X(Z)$ denotes the Pauli operator type, $i$ is a vertex index
satisfying $i\in A$ and $k$ represents the total number of vertex in the subsystem $A$. 
The component $g^{\ell,X(Z)}_{i}$ takes the binary values  $g^{\ell,X(Z)}_{i}=0$ or $1$.

By using the total $m$ truncated stabilizer generators $\{ g^A_{\ell}\}$, we construct the matrix $\tilde{K}$ given by \cite{sang2021}
\begin{eqnarray}
(\tilde{K}_A)_{\ell,\ell'}=
\begin{cases}
1 & \mbox{if}\:\: \{g^{A}_{\ell},g^{A}_{\ell'}\}=0\\
0 & \mbox{if}\:\: [ g^{A}_{\ell},g^{A}_{\ell'}]=0
\end{cases},
\label{K_rule}
\end{eqnarray}
where $\{g^{A}_{\ell},g^{A}_{\ell'}\}=0$ means that the truncated stabilizer generators $g^{A}_{\ell}$ and $g^{A}_{\ell'}$ are anti-commuting,
and $[g^{A}_{\ell},g^{A}_{\ell'}]=0$ means that the truncated stabilizer generators $g^{A}_{\ell}$ and $g^{A}_{\ell'}$ are commuting.
As a result, the matrix $\tilde{K}$ is the binary($Z_2$) $m \times m$ matrix. 

The rank of the matrix $\tilde{K}$ exhibits the number of anti-commutative pairs of the truncated stabilizer 
generators. 
Intuitively, the stabilizer generators touching the boundary of subsystem A are cut, and as a result,
some of them tend to anti-commute with other stabilizer generators. 
One then counts how many such instances occur. 
Apparently, this corresponds to counting the fundamental quantum correlation.

\subsection{Topological entanglement negativity}
Entanglement entropy in the pure topologically-ordered states is expected to exhibit a universal properties; 
having the universal constant named TEE~\cite{Levin2006,Kitaev2006}. 
That is, a recent work \cite{Fan_2024} conjectured that the negativity $\mathcal{N}_A$ also satisfies a scaling law such as
\begin{eqnarray}
\mathcal{N}_A = c|\partial A|- \gamma_N + \cdots,
\label{NA_scaling}
\end{eqnarray}
where $c$ is a non-universal constant and $|\partial A|$ is the perimeter of the subsystem A. 
(As explained later, this length unit is altered depending on the relevant degree of freedom of states.)
On the other hand, $\gamma_N$ is the TEN,  which can be a universal quantity in the sense that states belonging 
to an equivalent class of the (pure and/or mixed) topological order have the same value of TEN. 
The term $c|\partial A|$ in ${\cal N}_A$ means that the state exhibits area law of quantum entanglement.

The TEN can be extracted from negativity of subsystems by setting three adjacent subsystems, 
$A$, $B$ and $C$ such as
\begin{eqnarray}
\gamma_N &=& -\mathcal{N}_A - \mathcal{N}_B - \mathcal{N}_C - \mathcal{N}_{ABC}\nonumber\\
&+& \mathcal{N}_{AB} + \mathcal{N}_{BC} + \mathcal{N}_{AC}.
\label{def_TEN}
\end{eqnarray}
Throughout this work, for the value of TEN, the base of the logarithm is ``2'' since the negativity $\mathcal{N}_A$ is 
defined by using the same base of the logarithm.

\subsection{Exact value of TEN in the maximal decohered state $\mathcal{S}_D$}

In the previous section, we identified the modular theory from the gauging-out view point and 
by taking the quotient group, described by $\mathcal{A}_{D}/\{{\rm rX}\}$ of Eq.~(\ref{TC_modular_part}). 
The work \cite{Cai2026} has shown that the modular theory $\mathcal{M}$ obtained from the the nonmodular theory 
by factorizing the transparent sector gives a quantum dimension and the dimension is related to
the value of the TEN, given by
$$
\gamma_{N}\equiv \log_2 D_{\mathcal{M}}.
$$
where $D_{\mathcal{M}}$ is the quantum dimension of the modular theory $\mathcal{M}$.
We apply this observation to the preset case. 
Indeed, the modular theory $\mathcal{M}$ obtained by the transparent sector is just $\mathcal{A}_{D}/\{{\rm rX}\}$ of 
Eq.~(\ref{TC_modular_part}) and we have already seen that the modular theory is nothing but the TC type. 
Thus, in our target decohered state $\mathcal{S}_D$, 
the value of the TEN is
\begin{eqnarray}
\gamma_{N}\equiv \log_2 D_{\mathcal{A}_{D}/\{{\rm rX}\}}= 1.
\label{decoehred _exact_TEN}
\end{eqnarray}
This is our expectation value of the TEN of the color code under the maximal docoherence $\mathcal{E}^{XX}$.
In what follows, we further verify this value from different viewpoints.

\subsection{Analytical calculation of negativity for a small subsystem}
\begin{figure}[t]
\vspace{-0.5cm}
\begin{center}
\includegraphics[width=6.5cm]{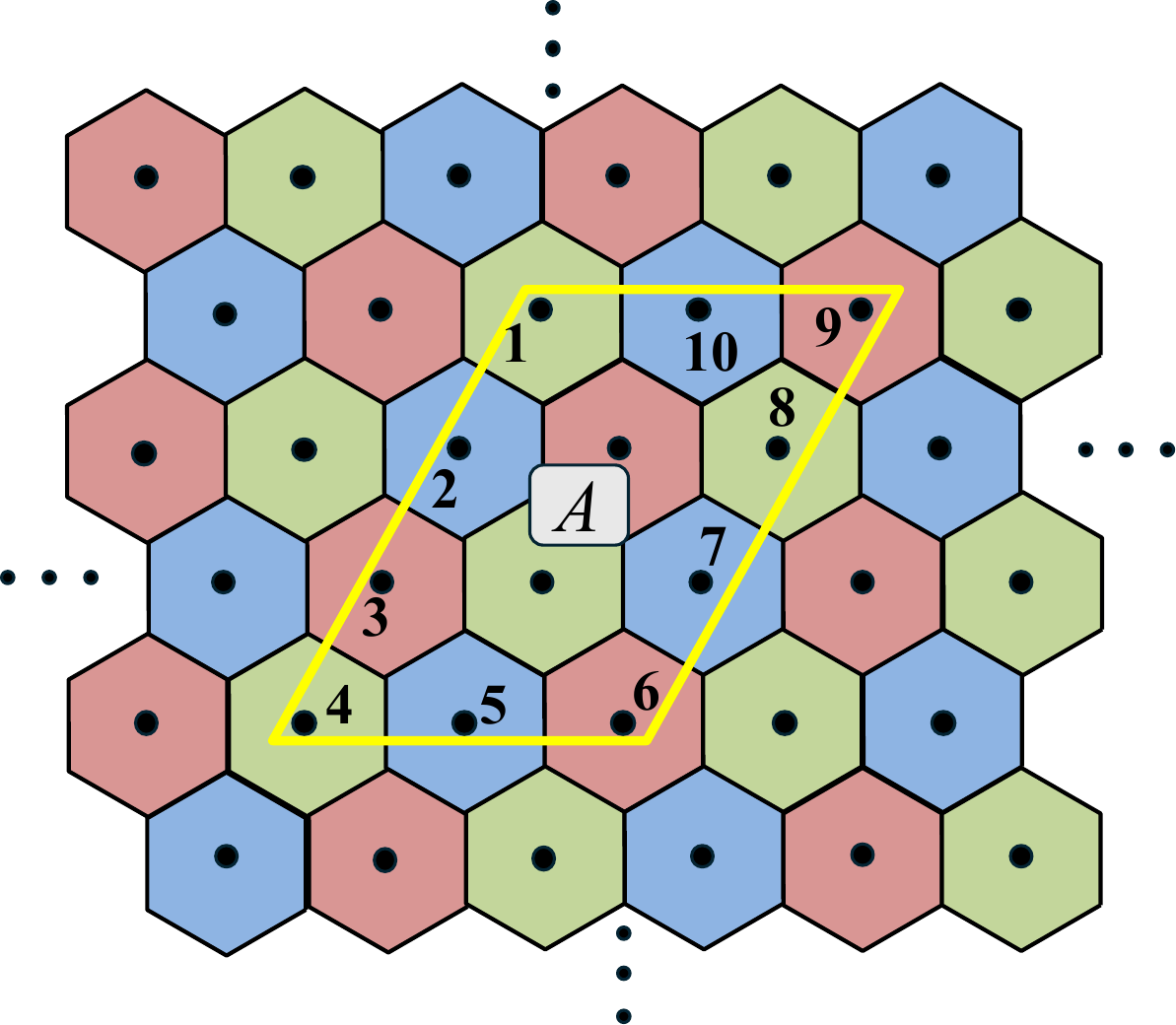}
\end{center}
\vspace{-0.5cm}
\caption{The color-code system for the analytical calculation of the negativity $\mathcal{N}_A$ without decoherences. The yellow parallelogram represents the boundary of the subsystem $A$. The total $20$ plaquette $Z$ and $X$ stabilizers are on the boundary. 
}
\label{Fig2}
\end{figure}

\begin{figure}[t]
\vspace{-0.5cm}
\begin{center}
\includegraphics[width=6.5cm]{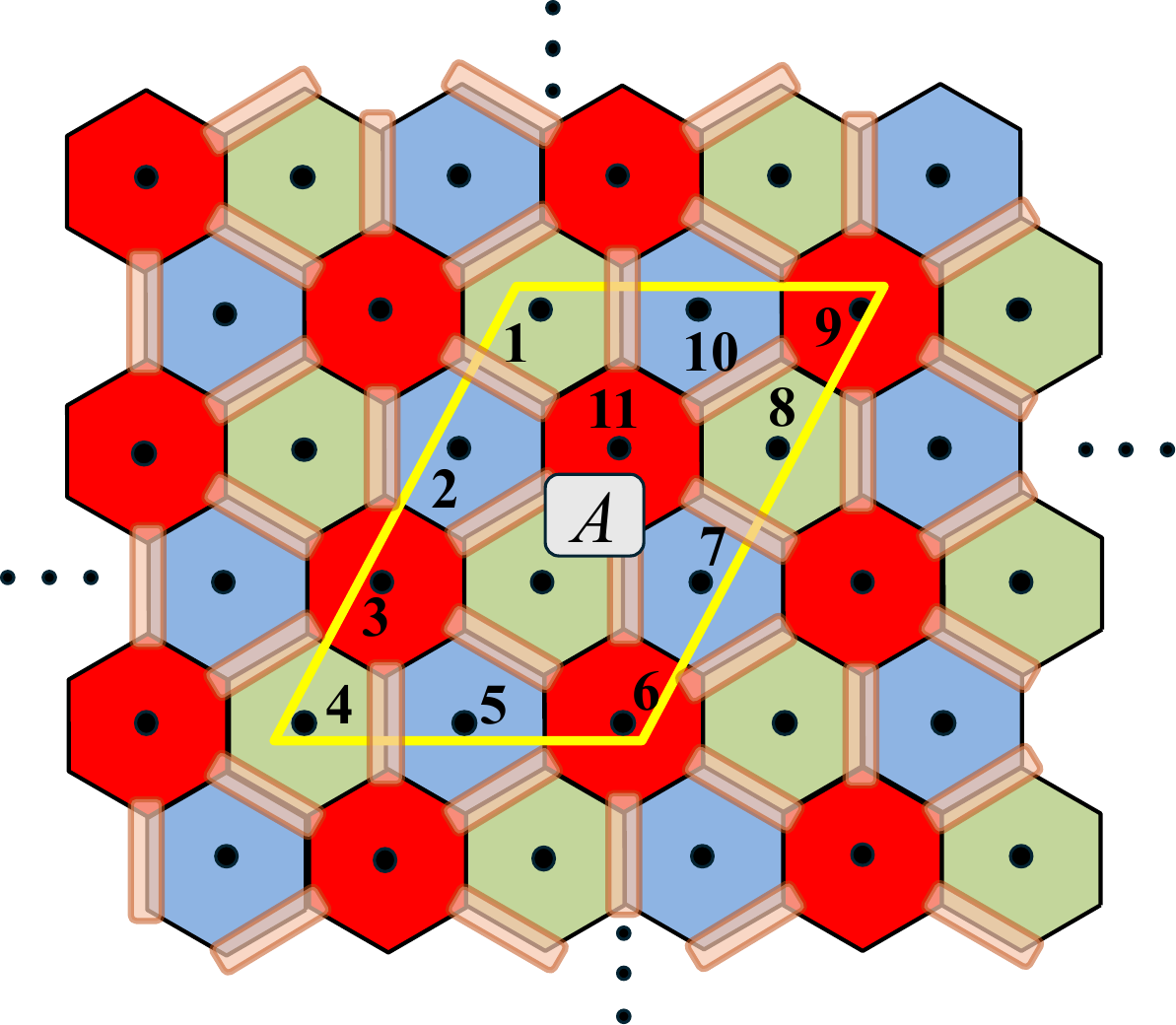}
\end{center}
\vspace{-0.5cm}
\caption{The color-code system for the analytical calculation of the negativity $\mathcal{N}_A$. We appliy the maximal decoherence $\mathcal{E}^{XX}$. The yellow parallelogram represents the boundary of the subsystem $A$. The elongated orange blocks on the red links represent the decoherence $\mathcal{E}^{XX}_{(v_r,v'_r)}$. In addition, the bright red plaquettes indicate that they are merged together into a single large stabilizer generator due to this orange decoherence $\mathcal{E}^{XX}$. 
}
\label{Fig3}
\end{figure}

By using the formula of Eq.~(\ref{N_A_formula}), we can calculate the negativity for both pure color code and the maximal decohered state by $\mathcal{E}^{XX}$. 
In this subsection, we calculate the negativity for the subsystem of the simple parallelogram 
as shown in Figs.~\ref{Fig2} and \ref{Fig3}. 
This choice of the subsystem seems to be conventional that has been used in some previous works 
for the calculation of the TEE for the honeycomb systems \cite{PhysRevResearch.6.L042063,Sriram_2023}.\\

\noindent \underline{Case I: Genuine color code state}\\
We calculate the concrete value of the negativity by using the formula of Eq.~(\ref{N_A_formula}). 
Here, we assume a very large system, but a small subsystem $A$ given by the yellow parallelogram 
as shown in Fig.~\ref{Fig2}. 
We first consider the genuine color code without decoherence represented by the set $\mathcal{S}_{\rm CC}$.
To this end, we remove four plaquette operators respecting the identities of Eqs.~(\ref{const2}) and ~(\ref{const3}).
We  chose plaquettes far from the subsystem $A$, meaning that the plaquettes $S^X_{(c,p)}$ and $S^Z_{(c,p)}$ surrounding 
the subsystem $A$ are all linearly-independent such that these are stabilizer generators. 
Based on this setup, we manipulate the form of $\tilde{K}_A$. 
Non-zero elements in the matrix $\tilde{K}_A$ come from $10\times 2=20$ plaquette generators $S^X_{(c,p)}$ and $S^Z_{(c,p)}$
on the boundary of the subsystem ($\partial A$) with label by the plaquette number, $1,2,\dots, 10$. 
Then, the matrix $\tilde{K}_A$ has a block-form such as (note that the following matrix is all $\mathbb{F}_2$(binary) matrix.)
\begin{eqnarray}
\tilde{K}_A=\left(
\begin{array}{cc}
\tilde{K}_{\partial A} & {\bf 0} \\
{\bf 0} & {\bf 0}
\end{array}
\right).
\end{eqnarray}
Here, $\tilde{K}_{\partial A}$ is $20\times 20$ matrix, the elements of which is constructed by $10\times 2=20$ plaquette generators 
$S^X_{(c,p)}$ and $S^Z_{(c,p)}$ on the boundary of the subsystem ($\partial A$). 
The matrix is further reduced to a  block-form as
\begin{eqnarray}
\tilde{K}_{\partial A}=\left(
\begin{array}{cc}
{\bf 0}_{10\times 10} & \tilde{K}^Z_{\partial A} \\
\tilde{K}^X_{\partial A} & {\bf 0}_{10\times 10}
\end{array}
\right),
\end{eqnarray}
where $\tilde{K}^{Z(X)}_{\partial A}$ are $10\times 10$ binary matrix.
By following the rule of Eq.~(\ref{K_rule}), the explicit form is given by
\begin{eqnarray}
&&\tilde{K}^{Z}_{\partial A}=
\begin{pmatrix}
0 & 1 & 0 & 0 & 0 & 0 & 0 & 0 & 0 & 1 \\
1 & 1 & 1 & 0 & 0 & 0 & 0 & 0 & 0 & 0 \\
0 & 1 & 1 & 1 & 0 & 0 & 0 & 0 & 0 & 0 \\
0 & 0 & 1 & 1 & 1 & 0 & 0 & 0 & 0 & 0 \\
0 & 0 & 0 & 1 & 1 & 1 & 0 & 0 & 0 & 0 \\
0 & 0 & 0 & 0 & 1 & 0 & 1 & 0 & 0 & 0 \\
0 & 0 & 0 & 0 & 0 & 1 & 1 & 1 & 0 & 0 \\
0 & 0 & 0 & 0 & 0 & 0 & 1 & 1 & 1 & 0 \\
0 & 0 & 0 & 0 & 0 & 0 & 0 & 1 & 1 & 1 \\
1 & 0 & 0 & 0 & 0 & 0 & 0 & 0 & 1 & 1
\end{pmatrix}.
\end{eqnarray}
We then obtain $\mathrm{rank}_{\mathbb{F}_2}\tilde{K}^{Z}_{\partial A}=8$. By using the same calculation, we also obtain $\mathrm{rank}_{\mathbb{F}_2}\tilde{K}^{X}_{\partial A}=8$. 
Thus, we get the value of the rank of $\tilde{K}_A$ and the negativity $\mathcal{N}_A$ as 
\begin{eqnarray}
\mathcal{N}_A&=&\frac{1}{2}\mathrm{rank}_{\mathbb{F}_2}\tilde{K}_A=\frac{1}{2}[\mathrm{rank}_{\mathbb{F}_2}\tilde{K}^{Z}_{\partial A}+\mathrm{rank}_{\mathbb{F}_2}\tilde{K}^{X}_{\partial A}]\nonumber\\
&=&8=10-2.
\label{N_A_color_code}
\end{eqnarray}
By using the scaling form of $\mathcal{N}_A$ of Eq.~(\ref{NA_scaling}), and supposing $c=1$ and $|\partial A|=10$ 
for the boundary length of the subsystem $A$, we obtain the value of TEN as $\gamma_N=2$. 
This value coincides with the value of TEE of the color code, which has been analytically obtained 
in \cite{Kargarian_2008,PhysRevResearch.6.L042063}.
Thus, the TEN of the color code without the decoherence is the same value of the TEE of the color code. 
In addition, the value of the TEN of the color code is double the value of TEN of TC since
that is $1$ as shown in \cite{Fan_2024,Wang_2025}.
\\

\noindent \underline{Case II: Color code under maximal $XX$ red-link decoherence}\\
We next consider the maximal decoherece case of $\mathcal{E}^{XX}$ ($p_r=1/2$) in the color code. 
The same assumptions to the former case are employed. 
Here, the docoherence $\mathcal{E}^{XX}$ changes the set $\mathcal{S}_{\rm CC}$ of original color code into the decohered one denoted by $\mathcal{S}'_D$ 
by following the decoherence update rule of the stabilizer formalism (See Appendix A). 
We slightly change the initial set of stabilizer generator of the color code for clear manipulation,
\begin{eqnarray}
\mathcal{S}'_{\rm CC}\longrightarrow\mathcal{S}'_{\rm CC}&=&\{S^X_{(r,p)}\}'+\{S^X_{(g,p)}\}'+\{S^X_{(b,p)}\}\nonumber\\
&+&\{S^Z_{(r,p)}\}+\{S^Z_{(g,p)}\}'+\{S^Z_{(b,p)}\}'.
\end{eqnarray}

Our aim is to calculate the negativity of the subsystem $A$ as shown in Fig.~\ref{Fig3}. 
Thus, we focus on the stabilizer generators residing on the boundary of the subsystem $A$, 
which give non-zero elements to the matrix $\tilde{K}_A$. 
With this observation and the update rule of the stabilizer formalism, we can pick up $18$ 
representative stabilizer generators from the set of $\mathcal{S}'_{D}$: (i) $10$-generators $S^X_{(c,p)}$ on the boundary $\partial A$. 
We denote this type of the stabilizers by $g^X_{k}$ ($k=1,2,\cdots, 10$), (ii) Total $7$-generators 
$S^Z_{(c,p)}$ ($(c,p)=1,2,4,5,7,8,10$) as shown in Fig.~(\ref{Fig3}). 
We denote this type of the stabilizers by $g^Z_{k}$ ($k=1,2,\cdots, 7$) 
(iii) In addition to the above stabilizers, there emerges a big stabilizer generator as a consequence of
the decoherence that is given as follows, 
$$
g^{ZB}=\cdots (S^Z_{(3)}S^{Z}_{(6)}S^Z_{(9)}S^{Z}_{(11)})\cdots  =\prod_{(r,p)}S^{Z}{(r,p)},
$$
which is the product of the bright red plaquettes in Fig.~\ref{Fig3}. 
This is created just by applying the decoherence $\mathcal{E}^{XX}$ ($p_r=1/2$) through the rule of the stabilizer formalism.

From these three types of the stabilizer generators, $\{g^X_{k}\}$, $\{g^Z_{k}\}$ and $g^{ZB}$ are truncated by
the boundary of the subsystem $A$, and the truncated stabilizer generators form some anti-commutative
pair among these truncated generators, thus giving non-zero elements in the matrix $\tilde{K}_A$. 
Then, we can analytically manipulate the matrix $\tilde{K}_A$. The matrix $\tilde{K}_A$ is the following block matrix 
\begin{eqnarray}
\tilde{K}_A=\left(
\begin{array}{cc}
\tilde{K}_{\partial A} & {\bf 0} \\
{\bf 0} & {\bf 0}
\end{array}
\right).
\end{eqnarray}
Here, $\tilde{K}_{\partial A}$ is $18\times 18$ matrix. 
The matrix $\tilde{K}_{\partial A}$ is further blocked-form such as
\begin{eqnarray}
\tilde{K}_{\partial A}=\left(
\begin{array}{cc}
\tilde{\tilde{K}}^Z_{\partial A} & {\bf 0}_{8\times 10}\\
{\bf 0}_{10\times 8} & \tilde{\tilde{K}}^X_{\partial A}
\end{array}
\right).
\end{eqnarray}
Here, $\tilde{\tilde{K}}^Z_{\partial A}$ is $10\times 8$ matrix each element of which is obtained by the observing the anti-commutation relation between $10$ generator 
$\{g^X_{k}\}$ and the total 8 generators $\{g^Z_k\}$ plus $g^{ZB}$. 
Also, $\tilde{\tilde{K}}^X_{\partial A}$ is $8\times 10$ matrix transposed by $\tilde{\tilde{K}}^Z_{\partial A}$.
As a result, we explicitly write down $\tilde{\tilde{K}}^Z_{\partial A}$ as
\begin{eqnarray}
&&\tilde{\tilde{K}}^Z_{\partial A}=
\begin{pmatrix}
0 & 1 & 0 & 0 & 0 & 0 & 1 & 0 \\
1 & 1 & 0 & 0 & 0 & 0 & 0 & 1 \\
0 & 1 & 0 & 1 & 0 & 0 & 0 & 1 \\
0 & 0 & 1 & 1 & 0 & 0 & 0 & 1 \\
0 & 0 & 1 & 1 & 0 & 0 & 0 & 1 \\
0 & 0 & 0 & 1 & 1 & 0 & 0 & 0 \\
0 & 0 & 0 & 0 & 1 & 1 & 0 & 1 \\
0 & 0 & 0 & 0 & 1 & 1 & 0 & 1 \\
0 & 0 & 0 & 0 & 0 & 1 & 1 & 1 \\
1 & 0 & 0 & 0 & 0 & 0 & 1 & 1
\end{pmatrix}.
\end{eqnarray}
Here, we obtain $\mathrm{rank}_{\mathbb{F}_2}(\tilde{\tilde{K}}^Z_{\partial A})=7$. 
By using the same manipulation, we obtain $\mathrm{rank}_{\mathbb{F}_2}(\tilde{\tilde{K}}^X_{\partial A})=7$.
As a result, the negativity $\mathcal{N}_A$ for the decohered state $\mathcal{S}'_D$ is observed as 
\begin{eqnarray}
\mathcal{N}_A&=&\frac{1}{2}\mathrm{rank}_{\mathbb{F}_2}\tilde{K}_A=\frac{1}{2}[\mathrm{rank}_{\mathbb{F}_2}\tilde{K}^{Z}_{\partial A}+\mathrm{rank}_{\mathbb{F}_2}\tilde{K}^{X}_{\partial A}]\nonumber\\
&=&7=8-1.
\label{N_A_decohered_color_code}
\end{eqnarray}
Here, if we employ the scaling form of $\mathcal{N}_A$ of Eq.~(\ref{NA_scaling}) as well as $c=1$ 
and also $\gamma_N=1$ (the TC value), then we have $|\partial A|=8$.
At first glance, this value of $|\partial A|$ is puzzling since the length of the boundary of the subsystem $A$ in 
Fig.~\ref{Fig3} is $10$ as used in the genuine color code case.
However, this discrepancy of the value of ${\cal N}_A$ is actually plausible in a sense, since under the maximal the red-link $XX$ decoherence, the system changes as 
\begin{eqnarray}
\text{color code} \simeq \text{toric code} \times \text{toric code} \longrightarrow 
\text{im toric code}. \label{change_imTO}
\end{eqnarray}
The surviving single TC resides on the emergent triangular lattice, sites of which are located on red 
plaquettes of the color code as seen in Fig.~\ref{Fig3}.
Figure \ref{Fig3} eloquently shows us that the employed subsystem $A$ {\it is not} commensurate
with the emergent triangular lattice. 
(In the pure state level, such a transformation of the lattices has been already suggested 
in~\cite{Kubica_2015,Haghighi2025}. 

\begin{figure}[t]
\vspace{-0.5cm}
\begin{center}
\includegraphics[width=6.5cm]{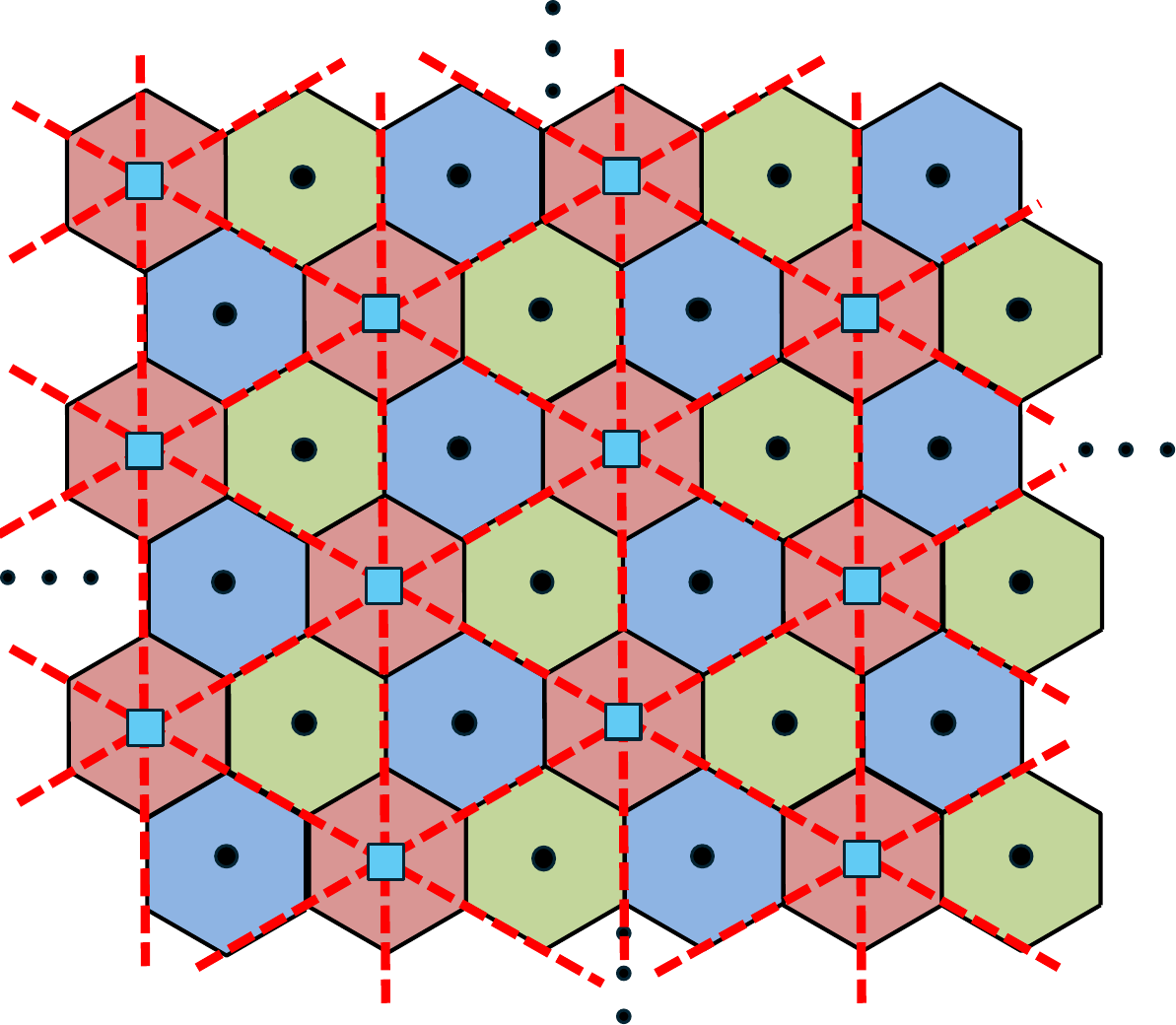}
\end{center}
\vspace{-0.5cm}
\caption{The emergent triangular lattice on the color-code system. 
The right-blue square symbols residing on the centers of the red plaquettes denote the sites of 
the triangular lattice. 
The links are represented by the red-dashed lines corresponding to the triangular TC lattice.} 
\label{Fig4}
\end{figure}

In Sec.~V, we numerically calculate ${\cal N}_A$ for subsystems with various shape and find that 
the values of ${\cal N}_A$ perfectly satisfy the scaling law Eq.~(\ref{NA_scaling}) for subsystems 
{\it commensurate with the triangular TC lattice} displayed in Fig.~\ref{Fig4}.
There, the length of the boundary of $A$, $|\partial A|$, is measured with unit of the lattice spacing of 
the emergent triangular lattice, giving the values $c=1$ and $\gamma_N=1$.
This fact indicates that a negativity ${\cal N}_A$ can be a good measure 
for the change of states such as Eq.~(\ref{change_imTO}), 
accompanying change of geometrical as well as topological properties as in the present case.
More details will be explained in Sec.~V after displaying the numerical results.

\section{Numerical calculations by using the efficient stabilizer algorithm}
In this section, we study the evolution of the system from the color code to the decohered state 
preserving the TC data by employing the efficient stabilizer algorithm \cite{gottesman1998,aaronson2004}. 
This numerical scheme let us examine the system's behavior in large-size systems. 
In the previous sections, we mainly investigated the maximal decoherence limit $p_r=1/2$ for $\mathcal{E}^{XX}$. 
Here, we employ stochastic decoherence, $\mathcal{E}^{XX}_{(v_r,v'_r)}$ on each red link $(v_r,v'_r)$ being applied with
probability $p$. 
We denote the obtained state by $\rho^s_{D}$ for $s$-th trajectory of the stochastic process. 
For the $p=1/2$ limit, $s$-dependence vanishes, and $\rho^s_D\longrightarrow \mathcal{E}^{XX}[\rho_{\rm CC}]$ for any $s$. 
By using the numerical methods, we verify the existence of the imTO through the observations of 
the negativity and TEN.
In addition to this, we investigate in detail how the stochastic system behaves for intermediate values of $p$. 

In this setup, we observe the ensemble average of the negativity $\mathcal{N}_A$ and TEN, given by
\begin{eqnarray}
\langle \mathcal{O}\rangle \equiv \mathbb{E}[\mathcal{O}]=\frac{1}{N_s}\sum_{s}\mathcal{O}^s, 
\label{average_observables}
\end{eqnarray}
where $\mathcal{O}^s$ represents the value of the observable $\mathcal{O}$ for $s$-th sample.
We take  $\mathcal{O}=\mathcal{N}_A$ and $\gamma_N$ and $N_s$ is the total number of trajectory samples and $N_s=\mathcal{O}(10^3)$ 
throughout this work. 

In the numerical calculation, we start with the color code state given by the set of the stabilizer 
generators $\mathcal{S}^{\rm CC}$ as the initial state. 
For the set $\mathcal{S}^{\rm CC}$, we perform the numerical update scheme corresponding to $\mathcal{E}^{XX}_{(v_r,v'_r)}$ with the probability $p$
as explained in Appendix A.

\subsection{Observation of TEN}
\begin{figure}[t]
\begin{center}
\includegraphics[width=5.5cm]{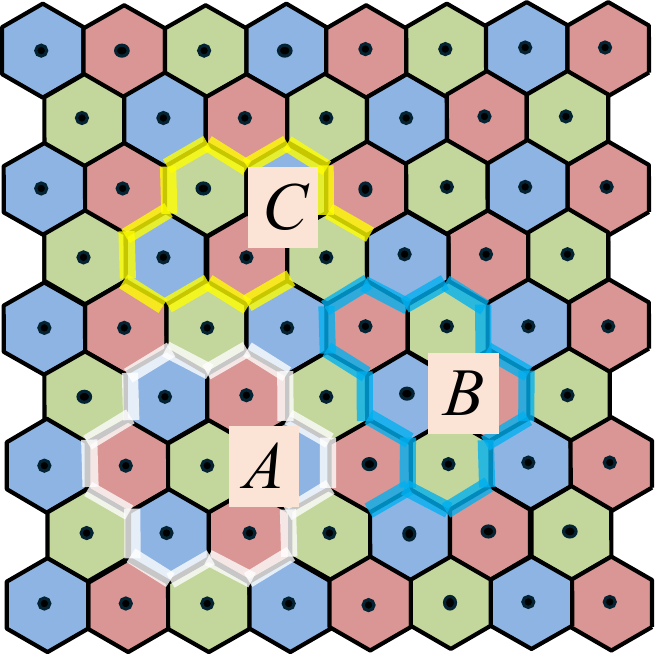}
\end{center}
\caption{Subsystem complex used for calculation of the TEN. 
Boundary vertices for each subsystem reside on the sites on the boundary lines of
the honeycomb lattice colored with while, blue and yellow.
Three boundaries do not share any sites in order to avoid overlaps.
Displayed subsystems are the smallest ones for the calculation.
We expand those by keeping their hexagonal shapes composed of centers of boundary plaquettes
for constructing larger subsystems. 
The $A$-subsystem contains seven plaquettes. Representatively, we call the shape of the subsystems for the calculation of TEN 7-plaquette.
}
\label{Fig_TEN_partition}
\end{figure}

We study the TEN, $\langle \gamma_N\rangle$, as a function of $p$. 
For the Practical calculation, we use specific shapes of the subsystems for the estimation of the TEN. 
The shapes of the subsystems $A$, $B$ and $C$ are shown in Fig.~\ref{Fig_TEN_partition}
for the smallest configuration, in which subsystems have seven, six and five plaquttes, respectively. 
We call this shape of the subsystems for the calculation of the TEN ``7-plaquette'' through
the number of the plaquette of the $A$-subsystem.
We choose this setup of the subsystems since our previous calculation of the topological entanglement
entropy (TEE) in the triangular TC \cite{Kataoka_2026} indicates that a hexagon is suitable for the calculation
of the topological quantities as it contains only obtuse angles.
However, we verified the stability of the TEN by considering various shapes of subsystem complex, some of which are shown in Appendix D. 

\begin{figure}[t]
\begin{center}
\includegraphics[width=8.5cm]{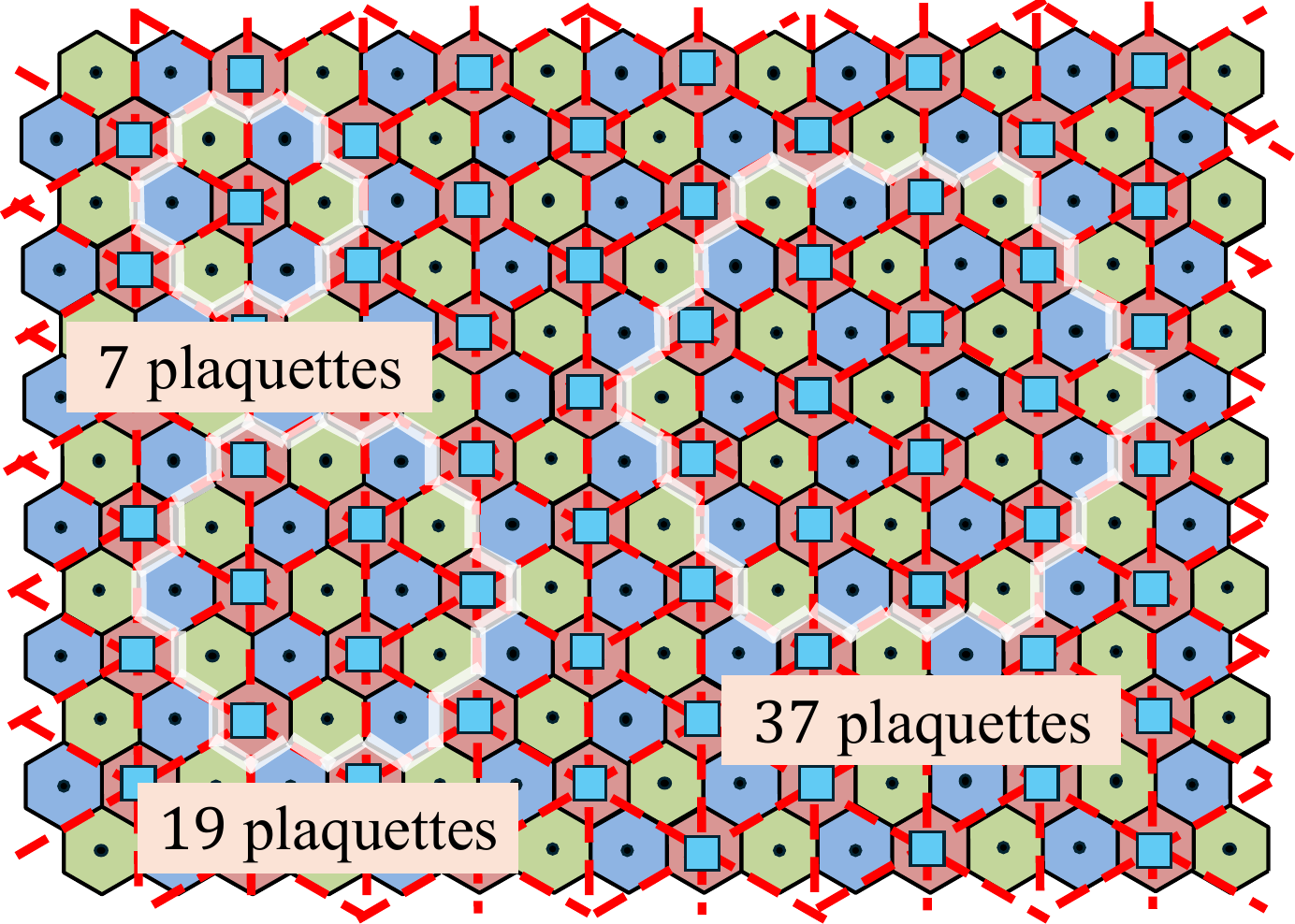}
\end{center}
\caption{Three kinds of $A$-subsystems used for calculation of the TEN,
represented by hexagonal region.
These are consisting of 7 plaquettes, 19 plaquettes and 37 plaquettes, respectively. 
For each choice of the subsystem $A$, the subsystems $B$ and $C$ are accompanied with it 
in the same manner as in Fig.~\ref{Fig_TEN_partition}.
}
\label{Fig_plaquettes}
\end{figure}

We perform calculation for cases with subsystem sizes larger than the 7-plaquette configuration. 
We define the subsystems $A$, $B$, and $C$ similarly to the 7-plaquette configuration, then
we calculate the TEN for three kinds of setups in which subsystem $A$ is given 
by the hexagonal region consisting of 19 plaquettes and 37 plaquettes, as shown in Fig.~\ref{Fig_plaquettes}.
Through them, we investigate the system-size dependence of the TEN and its variance.


\begin{figure}[t]
\begin{center}
\includegraphics[width=7cm]{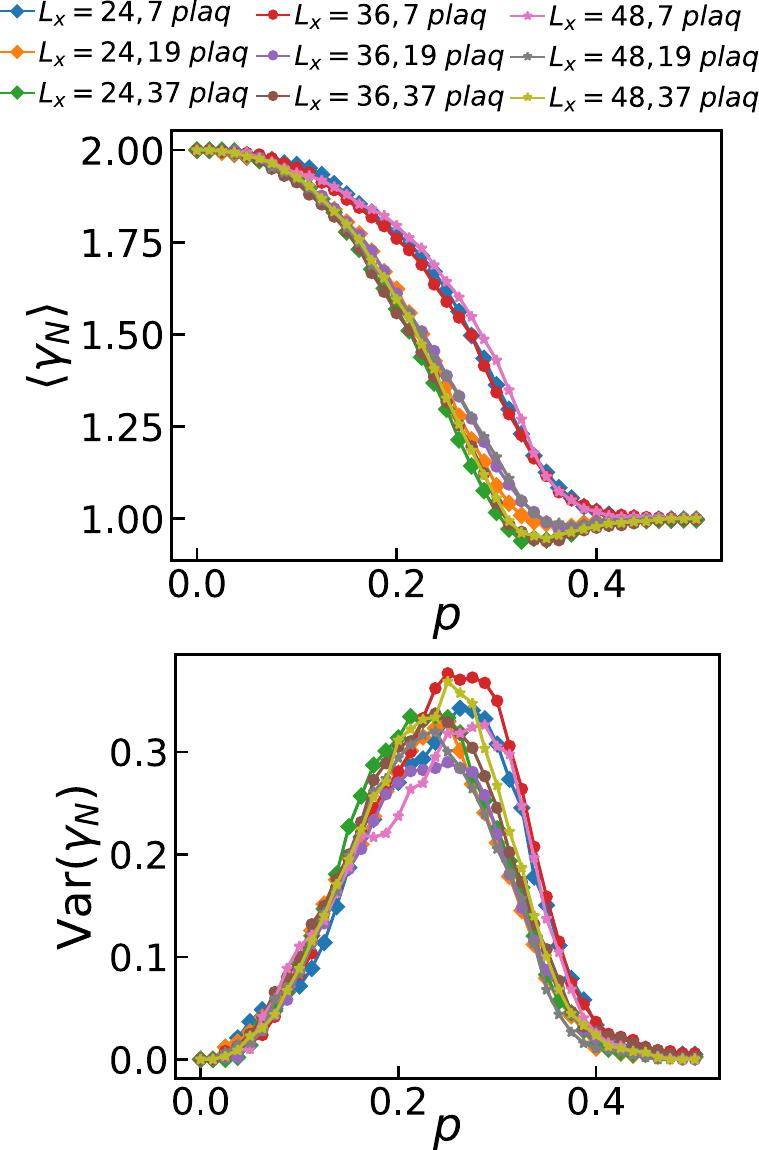}
\end{center}
\caption{(Upper panel) The value of TEN, $\langle\gamma_N\rangle$, as a function of $p$.
The data of all the sizes exhibit a smooth change from the color code value $2$ (in unit $\ln 2$)
to that of the TC, unity.
(Lower panel) Variance of the TEN. 
The data show a rather large peak from $p\sim 0.1$ to $0.4$ without clear subsystem-size dependence.
In both panels, we plot  the data for the 7-plaquette, 19-plaquette, and 37-plaquette configurations
explained in Fig.~\ref{Fig_plaquettes} and we set $L_x=L_y=24,36$ and $48$.
The number of samples is $\mathcal{O}(10^3)$. 
}
\label{Fig_TEN}
\end{figure}

Figure~\ref{Fig_TEN} displays the obtained numerical data of the TEN and its variance
for 7-plaquette, 19-plaquette and 37-plaquette configurations with the system size 
$L_x=L_y= 24, 36$ and $48$.

The $\langle \gamma_N\rangle$ of all the sizes exhibit a smooth change from the color code value 2
to that of the TC, unity.
Furthermore, no system-size dependence of the TEN is observed among them.
This behavior implies that the change of the system is not a sharp phase transition but 
a crossover like one. 

Behavior of the variance of $\langle \gamma_N\rangle$ provides us more information about the criticality of the system.
The data in ~Fig.~\ref{Fig_TEN} show that it has a large and broad peak between $p\sim 0.1$ and $p\sim 0.4$.
This result may support a crossover in that region, but the peak is too large and also its shape
is somewhat peculiar (i.e., a simple triangle form)
from the view point of the ordinary crossover or higher-order phase transition 
such as the Kosterlitz-Thouless transition in the two-dimensional XY model~\cite{Kosterlitz_1973,Janke_1993,KSKIM_2017}.

In fact, the above behavior of the TEN is in sharp contrast to that of the TEE in the TC on the triangular
lattice under measurement with outcome recoding~\cite{Kataoka_2026}. 
The TEE exhibits clear system-size dependence, and its variance has a sharp peak.
Finite system-size scaling analysis works quite efficiently for them and the critical exponents are obtained. 

\subsection{Observation of $\mathcal{N}_A$}

In this subsection, we numerically investigate the negativity, ${\cal N}_A$, in detail.
Motivation of this study is two fold;
(1) To see how the scaling law in Eq.~(\ref{NA_scaling}) is satisfied in the present stochastic
decoherence process.
In fact in Sec.~IV.D, we have already remarked that the scaling law does not hold for an
arbitrary subsystem $A$, and commensurability with the emergent triangular lattice 
is a key point for the law.
(2) In the previous subsection, we observed some peculiar behavior of the TEN and its variance.
We wonder if large fluctuations in the TEN come from those of the ingredient negativity or
the manipulation to obtain the TEN by combining the negativity of the subsystems.
Following numerical study clarifies both of the above issues.

\begin{figure}[t]
\begin{center}
\includegraphics[width=8.5cm]{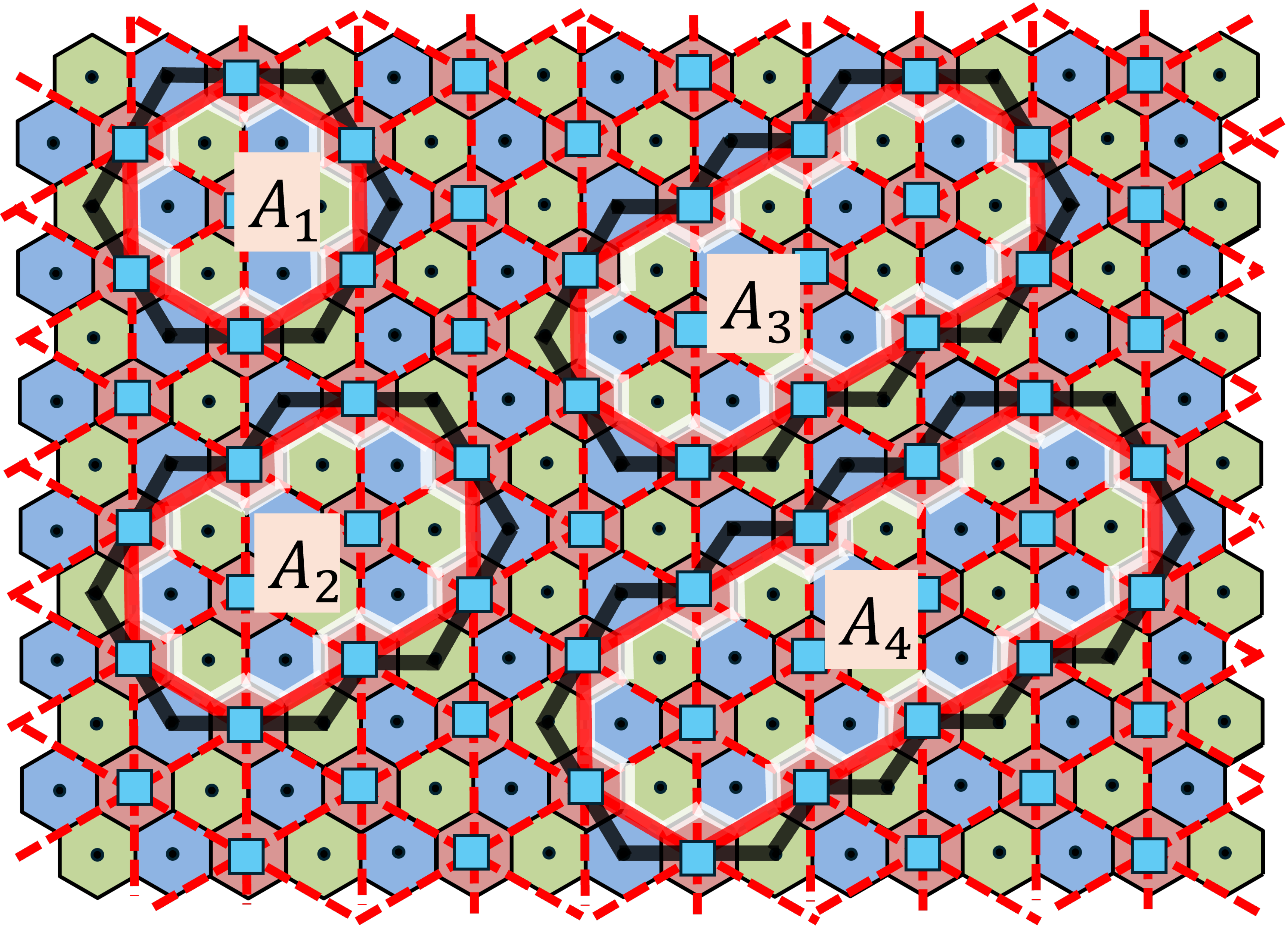}
\end{center}
\caption{Setting of the subsystems for the calculation of $\mathcal{N}_{A}$;
Subsystems $A_1$, $A_2$, $A_3$ and $A_4$ in the color code honeycomb lattice.
Zigzag lines (the black-solid lines) surrounding the subsystems represent boundaries 
and their lengths are used for the scaling law
Eq.~(\ref{NA_scaling}) for the color code, and practically, the boundary length $|\partial A_{i}|=12$, $16$, $20$ and $24$, respectively. The scaling law gives $c=1$ and $\gamma_N=2$.
Straight lines (the red-solid lines) represent the boundaries of subsystems $A_1\sim A_4$ from
the triangular-lattice view.
 From the perspective of the emergent triangular lattice in the TC limit, sites of which are located on the 
 red plaquettes as in Fig.~\ref{Fig4}, each subsystem can be regarded as a subsystem on the triangular lattice.
 Boundary length $|\partial A|_{\rm TC}$ is measure with the unit of  the lattice spacing of the triangular lattice
 (the distance between centers of two adjacent red palquettes),
 and then, $|\partial A_i|_{\rm TC}=6, 8, 10$ and $12$, respectively.
 With these values, the scaling law of ${\cal N}_A$, Eq.~(\ref{NA_scaling}), is satisfied with $c=1$ and 
 $\gamma_{ N}=1$, precisely as in the TC.
}
\label{Fig_subsystem_NA}
\end{figure}

In Fig.~\ref{Fig_subsystem_NA}, we display subsystems for which the negativity is calculated.
All subsystems on the honeycomb lattice are commensurate with the triangular lattice
and therefore, they are subsystems on the triangular lattice.
These subsystems have two kinds of perimeter length, one on the honeycomb and the other
on the triangular lattices.
If the negativity ${\cal N}_A$ is a good measure for observing the evolution from the color code
to TC, its scaling law Eq.~(\ref{NA_scaling}) is satisfied by suitably manipulation the definition
of the perimeter length from $|\partial A|$ (on the honeycomb) to $|\partial A|_{\rm TC}$ (on the triangular).

\begin{figure}[t]
\begin{center}
\includegraphics[width=6.5cm]{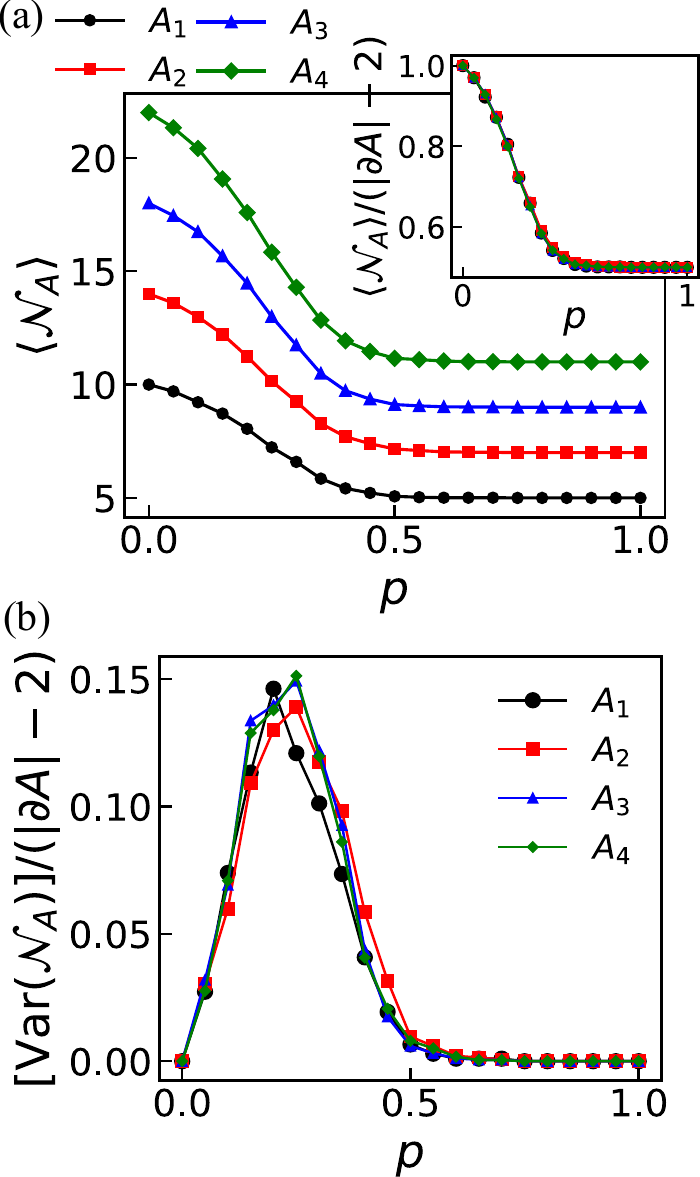}
\end{center}
\caption{(a) Calculations of $\langle \mathcal{N}_A\rangle$ as a function of $p$.
Inset: $\langle \mathcal{N}_A\rangle$ is normalized divided by $(|\partial A_i|-2)$ as in the color code, ${\cal N}_A= |\partial A|-2$.
It is also expected that for the TC limit, 
${\cal N}_A= |\partial A|_{\rm TC}-1={1\over 2}(|\partial A|-2)$ suggesting this normalization.
Obtained numerical data of $\langle \mathcal{N}_A\rangle$ obviously support this scaling law. 
(b) Sample-to-sample variance of $\mathcal{N}_A$ divided by ($|\partial A_i|-2$).
}
\label{Fig_NA}
\end{figure}

Based on the above observation, we display the data of the negativity in Fig.~\ref{Fig_NA}(a) and 
the normalized ones with $(|\partial A|-2)$ in the inset of Fig.~\ref{Fig_NA}(a).
We find that the excellent scaling holds.
See Fig.~\ref{Fig_NA}(a).
The data obviously show that the system changes smoothly from the color code with 
$\gamma_N=2$ to the TC with $\gamma_N=1$, and the scaling coefficient is universally $c=1$
as a result of changing the unit of length.

We have also performed similar calculations for subsystems with distinct shapes incommensurate
with the triangular lattice and found that a similar scaling to the above does not hold. 
Appendix D shows some of examples.

Let us turn to the second motivation.
Figure ~\ref{Fig_NA}(b) shows the variance of the negativity $\langle{\cal N}_A\rangle$ divided by $(|\partial A_i|-2)$
for $A_i (i=1\sim 4)$.
We find that the negativity has a large fluctuation as the TEN in the preceding subsection,
and conclude that the large fluctuations of the topological order exist in each subsystem.
The TEN as well as the negativity display the genuine properties of the intermediate states
between the two topologically-ordered states, the color code and the TC.

The large sample-to-sample fluctuations observed in the above implies that in the intermediate
regime, {\it there are so many variety of stabilizer configurations from the viewpoint of the negativity ${\cal N}_A$
coming from its genuine definition}.
This property may be specific for the present system or rather universal for the process from
one topological state to another one.
This problem is very interesting, but currently we have no practical methods to analyzing it.
We leave it for a future problem.

\section{Conclusion and discussion} 

In this work, we studied the color code under the decoherence of the $XX$ type acting on
red edges in a stochastic manner, and observed how the decohered color code changes to the mixed state. 
The mixed state is the imTO described by nonmodular theory, which has no counterpart in pure-state
ground states of a local Hamiltonian.
Numerical studies of the negativity and the TEN by using the stabilizer formalism played an essentially
important role to observe how the color code changes to the topological mixed state, the TC, 
under the decoherence.

Based on the stabilizer formalism, we elucidated the properties of the imTO by using 
the gauging out perspective and showed that it is a nonmodular theory, in which transparent anyons exist. 
We further investigated the anyon data of the decohered color code, and found that the modular theory 
exhibits the anyon data of a single TC type. 
We concluded that the decohered color code state is the imTO and we clarified its characterization 
by the TEN, which has not been reported so far. 
This study suggests that the imTO is a ubiquitous mixed state, given by the topological code under various decoherence conditions. 

In addition, we clearly elucidated the symmetry aspect beyond the on-site symmetry, that is, we discussed the emergent strong 1-form symmetries that the imTO possesses in the decohered color code. 

We next showed that from the modular theory obtained by factorization,  
the decohered color code can be characterized by a universal value of the negativity, i.e., the TEN,
as recently proposed in Ref.~\cite{Cai2026}. 
As a first step, by the analytical calculation of the negativity for a small system, 
we showed that for the genuine color code, the value of the negativity indicates that the TEN is $2\ln 2$, 
while in the limit of the maximal decoherence, the negativity takes a different value.

In the last part of this study, we carried out large-scale numerical simulations by using 
the efficient stabilizer algorithm, where the decoherence was set as a stochastic one. 
Numerically, we found that the TEN clearly changes from $2\ln2$ to $\ln 2$ as increasing 
the probability $p$ of the $XX$-type decoherence. 
The obtained data showed that the TEN varies as expected from $2\ln2$ to $\ln 2$, 
but its smooth behavior, as well as the system-size independent variance, implies that the change is of
cross-over type, not a sharp-mixed phase transition.
We leave this point as an interesting future problem.

As another future work, 
it is important to study how other topological codes exhibit imTO's under certain decoherence and 
examine whether the observed phenomena in this work are universal or specific to the present target system.
In particular,  by employing some other numerical methods such as tensor network~\cite{Darmawan_2017},
complementary and deep understanding of the evolution of the topological states under decoherence 
can be obtained.
As another aspect of the state evolution, 
under the change from the color code to the imTO, how the mixed state ``gap'', which can be efficiently estimated by the conditional mutual information and its Markov length~\cite{Sang_PRL_2025,negari2025,Zhang_2025}, behaves is an important problem.

\section*{Data availability}
The data that support the findings of this study are available from the authors upon reasonable request.\\

\bigskip
{\it Acknowledgements.---}
This work is supported by JSPS KAKENHI: JP23K13026 and JP26K06956(Y.K.) and JP23KJ0360 and JP26K17056(T.O.). 
The computations in
this work were done using the facilities of the Supercomputer Center, The Institute for Solid State Physics,
The University of Tokyo (2026-Ba-0068).


\renewcommand{\thesection}{A\arabic{section}} 
\renewcommand{\theequation}{A\arabic{equation}}
\renewcommand{\thefigure}{A\arabic{figure}}
\setcounter{equation}{0}
\setcounter{figure}{0}
\appendix

\setcounter{section}{0}
\renewcommand{\thesection}{\Alph{section}}
\makeatletter
\renewcommand{\theHsection}{\Alph{section}}
\makeatother



\makeatletter
\renewcommand{\theHfigure}{\thesection.\arabic{figure}}
\renewcommand{\theHtable}{\thesection.\arabic{table}}
\renewcommand{\theHequation}{\thesection.\arabic{equation}}
\makeatother

\section{Update rule of local decoherence in stabilizer formalism}
Here, we consider a local decoherence $\mathcal{E}^{XX}_{(v_r,v'_r)}$ mainly for $p_r=1/2$ (fixed point) corresponding to the
measurement with the red-link $X_{v_r}X_{v'_r}$ operator without recording outcomes, that is, the local maximal 
decoherence. 
The outcome denoted by $\beta_{(v_r,v'_r)}$ is $\beta_{(v_r,v'_r)}=\pm 1$ since $(X_{v_r}X_{v'_r})^2=I$, and then,
the channel of $\mathcal{E}^{XX}_{(v_r,v'_r)}$ for $p_r=1/2$ is given as \cite{weinstein2022,Dias_2023}
\begin{eqnarray}
\mathcal{E}^{XX}_{(v_r,v'_r)}[\rho]=\sum_{\beta_{(v_r,v'_r)}=\pm}P^{X_{v_r}X_{v'_r}}_{\beta_{(v_r,v'_r)}}\rho P^{X_{v_r}X_{v'_r}\dagger}_{\beta_{(v_r,v'_r)}},
\label{local_decoherence_channel}
\end{eqnarray}
where $P^{X_{v_r}X_{v'_r}}_{\beta_{(v_r,v'_r)}}$ is a projection operator of $X_{v_r}X_{v'_r}$ with outcome $\beta_{(v_r,v'_r)}$, 
$\displaystyle{P^{X_{v_r}X_{v'_r}}_{\beta_{(v_r,v'_r)}}=\frac{1}{2}[I+\beta_{(v_r,v'_r)}X_{v_r}X_{v'_r}]}$.

We observe how the local decoherence channel $\mathcal{E}^{XX}_{(v_r,v'_r)}$ acts on a state $\rho$ practically.  
The density matrix $\rho$ is represented by $k$-stabilizer generators $\{g_{\ell}\}$ with $\ell=0,1,2,\cdots, k-1$ and
the total qubit number is set $N_q$,  
\begin{eqnarray}
\rho=\frac{1}{2^{N_q-k}}\prod^{k-1}_{\ell=0}\frac{I+g_{\ell}}{2}.
\label{dens_stab}
\end{eqnarray}
As explained in Refs.\cite{weinstein2022}, the introduction of the local decoherence channel $\mathcal{E}^{XX}_{(v_r,v'_r)}$ is 
efficiently implemented in the stabilizer algorithm.

By applying $\mathcal{E}^{XX}_{(v_r,v'_r)}$ to $\rho$, the density matrix represented by the stabilizer generators changes as
\begin{eqnarray}
\mathcal{E}^{XX}_{(v_r,v'_r)}[\rho]&=&\sum_{\beta_{(v_r,v'_r)}=\pm}P^{X_{v_r}X_{v'_r}}_{\beta_{(v_r,v'_r)}}\rho P^{X_{v_r}X_{v'_r}\dagger}_{\beta_{(v_r,v'_r)}}\nonumber\\
&=&\biggr(\sum_{\beta_{(v_r,v'_r)=\pm}}P^{X_{v_r}X_{v'_r}}_{\beta_{(v_r,v'_r)}}\biggl[\frac{1+\tilde{g}_{0}}{2}\biggr]P^{X_{v_r}X_{v'_r}\dagger}_{\beta_{(v_r,v'_r)}}\biggr)\nonumber\\&&\times \biggl[\frac{1}{2^{N_q-k}}\prod^{k-1}_{\ell=1}\frac{1+\tilde{g}_{\ell}}{2}\biggr]\nonumber\\
&=&\frac{1}{2^{N_q-k+1}}\prod^{k-1}_{\ell=1}\frac{1+\tilde{g}_{\ell}}{2},
\end{eqnarray}
where on the second line, we have performed a standard transformation between the stabilizer 
generators \cite{Nielsen2011}, i.e., the set of stabilizer generators $\{g_\ell\}$ 
is changed onto the one denoted by $\{\tilde{g}_\ell\}$, in which at most one stabilizer generator labeled 
by $\tilde{g}_{0}$ anti-commute with $X_{v_r}X_{v'_r}$. 
Thus, application of the local decoherence $\mathcal{E}^{XX}_{(v_r,v'_r)}$ eliminates one stabilizer generator from the set of 
previous stabilizer generators, leading to the enhancement of the mixing of the state. 
This can be observed from the fact that the factor of the density matrix gets smaller 
as $\frac{1}{2^{N_q-k}}\longrightarrow \frac{1}{2^{N_q-k+1}}$ through the channel $\mathcal{E}^{XX}_{(v_r,v'_r)}$.

\section{$1$-form strong and weak symmetries}
On considering mixed states, two types of symmetries is to be introduced for 
the states and also for quantum channels applying to the states, namely, strong and weak 
symmetries \cite{Buca_2012,Albert2014,groot2022}.
In this work, we focus on the $Z_2$ 1-form symmetry, the generator of which is given by 
a loop operator $W_{Z_2}$ satisfying $W^2_{Z_2}=\hat{1}$, that is, the group is denoted as $\{\hat{1},W_{Z_2}\}$. 

First, we give the general definition of the 1-form strong symmetry expressed in terms of  density matrix,
$$
W_{Z_2}\rho=e^{i\theta}\rho,\;\;\; \rho W^{\dagger}_{Z_2}=e^{-i\theta}\rho,
$$
where $\rho$ is a density matrix and $\theta$ is a global phase factor. 
We next give the definition of the 1-form weak symmetry. It is  expressed as
$$
W_{Z_2}\rho W^\dagger_{Z_2} = \rho.
$$ The symmetry is satisfied in ensemble average level \cite{ma2024}.

The notion of the strong and weak symmetries are further defined on quantum channels. 
Generally, the quantum channel including decoherence is described by the Kraus-operator-sum 
form \cite{Nielsen2011,lidar2020},
$$
\mathcal{E}(\rho)\equiv\sum^{N-1}_{\ell=0}K_{\ell} \rho K^\dagger_{\ell},
$$ where $\{K_{\ell}\}$ are a set of Kraus operators satisfying $\sum^{N-1}_{\ell=0} K^\dagger_{\ell} K_{\ell}=\hat{I}$ with $\hat{I}$ being the identity 
operator, preserving $\mathrm{Tr}[\mathcal{E}(\rho)]=1$. 
The quantum channel $\mathcal{E}$ induces changes in mixed states. Generally, it is not invertible. 
Under a condition, there is a recovery channel \cite{Petz1988,Junge2018}. 
We here give the definition of $Z_2$ 1-form strong symmetry for the channel as 
$$
K_{\ell}W_{Z_2}=e^{i\theta} W_{Z_2} K_{\ell}
$$ 
for any $\ell$. 
On the other hand, the definition of the 1-form weak symmetry for the channel is expressed as 
$$
W_{Z_2}\biggl[\sum_{\ell}K_{\ell} \rho K^\dagger_{\ell}\biggr]W^\dagger_{Z_2}=\mathcal{E}(\rho).
$$
This definition does not require that each Kraus operator commutes with the non-trivial loop generator $W_{Z_2}$.

\begin{figure}[t]
\begin{center}
\includegraphics[width=8.8cm]{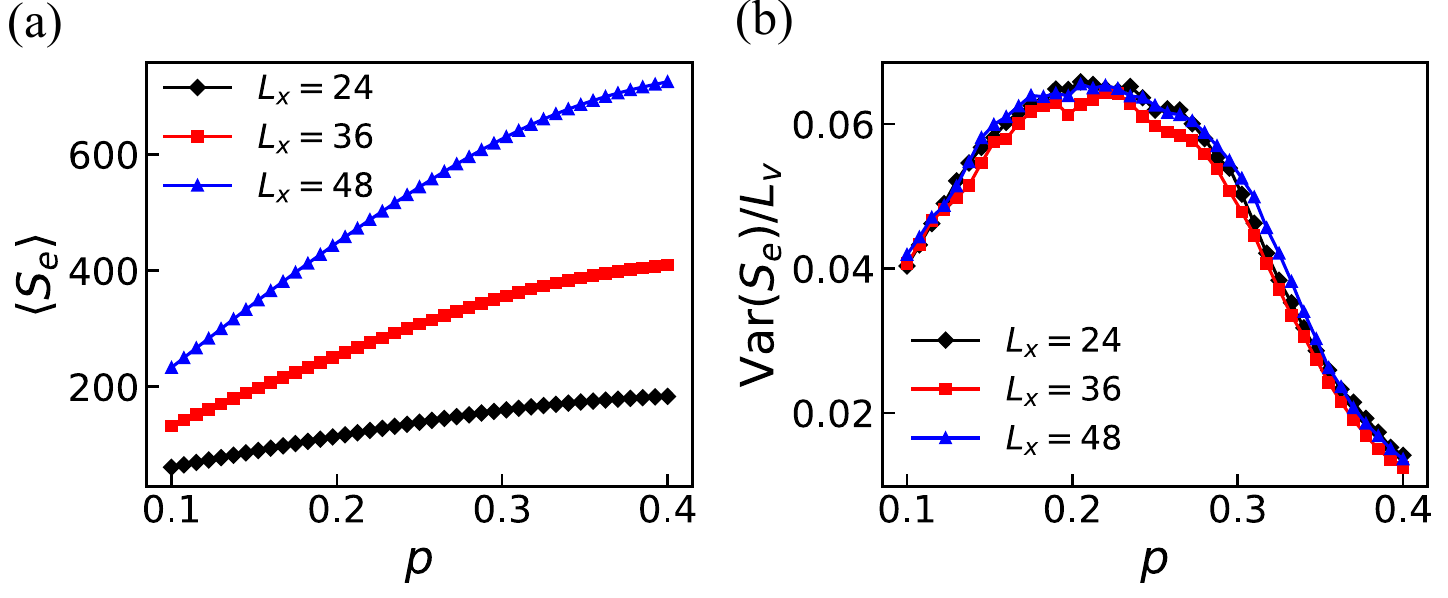}
\end{center}
\caption{$p$-dependence of $\langle S_e\rangle$[(a)] and its variance with volume scaled [(b)]. 
The number of samples is $\mathcal{O}(10^3)$.
}
\label{Fig_purity}
\end{figure}
\section{Behavior of purity in stabilizer numerics}
\setcounter{equation}{3}
In this Appendix, we show the additional numerical data exhibiting mixing behavior in the color code
as varying the probability $p$ in the stochastic decohered system. 
This result shows how entangled to environment the color code under the decoherence $\mathcal{E}^{XX}_{(v_r,v'_r)}$. 
To this end, we observe the logarithmic purity, denoted as $S_e$
\begin{equation}
S_e=-\text{log}_2\ \mathrm{tr}[\rho^2].
\label{purity}
\end{equation}
It measures the correlation between the system and environment, namely system-environment 
entanglement\cite{Ashida_2024,KOI2026_SEE}.

Numerically, by the efficient stabilizer algorithm, $S_e$ can be easily calculated: From the number of the 
independent stabilizer generators denoted by $N_{gi}$ of a state $\rho$, $S_e=N_v-N_{gi}$, where $N_{gi}$ is obtained 
by the set of the stabilizer generators $\mathcal{S}_D$. 

Figure ~\ref{Fig_purity} (a) is $p$-dependence of the average value of $S_e$, $\langle S_e \rangle$, where we used $\mathcal{O}(10^3)$ samples. 
$L_v=L_x\times L_y$. 
Moreover, we plot the rescaled variance obtained by many samples in Fig.~\ref{Fig_purity} (b). 
We observe that the variance exhibits behavior similar to ones of the negativity and TEN. 
The value of $\langle S_e \rangle$ is a monotonically decreasing function of $p$, as expected. 
The errors for each data point is much smaller compared to the original values but, 
we observe that a pronounced enhancement of the variance is observed. 
Moreover, as shown in Fig. ~\ref{Fig_purity} (b), when the variance is divided by the system volume, 
it exhibits an approximately size-independent behavior, and no divergence is observed.


\begin{figure}[t]
\begin{center}
\vspace{1 cm}
\includegraphics[width=8cm]{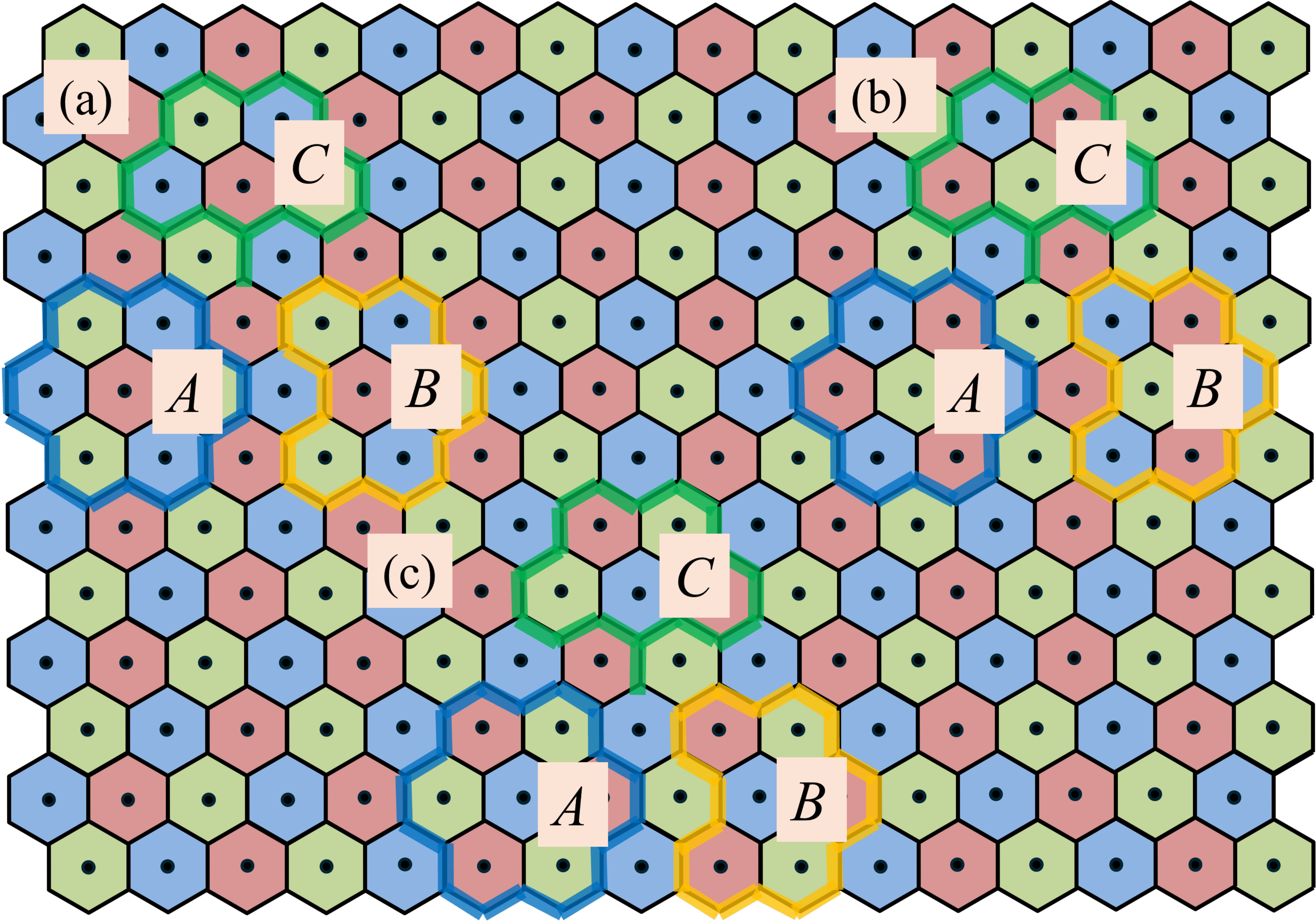}
\end{center}
\caption{Subsystem combinations (a), (b) and (c) for the negativity and TEN 
in the decohered color code. 
Color of the central plaquette in the subsystems in (a), (b) and (c) cases is red, green and blue, respectively.}
\label{Fig_TEN_partition2}
\end{figure}

\section{Stability of topological entanglement negativity}

In the main text, we remarked that the value of the negativity ${\cal N}_A$ generally depends on the shape of 
the subsystems except for the genuine color code.
In fact, in this Appendix, we show the numerical data of the negativity $\mathcal{N}_A$ for the subsystems
displayed in Fig.~\ref{Fig_TEN_partition2}. 
There, only the case (a) is commensurate with the triangular lattice (as shown in Fig.~\ref{Fig4} in the main text) and can be regarded as a subsystem from the view of the triangular lattice. 
This case should be compared with the subsystems in Fig.~\ref{Fig_subsystem_NA} in the main text.
The subsystems (b) and (c) are incommensurate with the triangular lattice.

\begin{figure*}[t]
\begin{center}
\includegraphics[width=16cm]{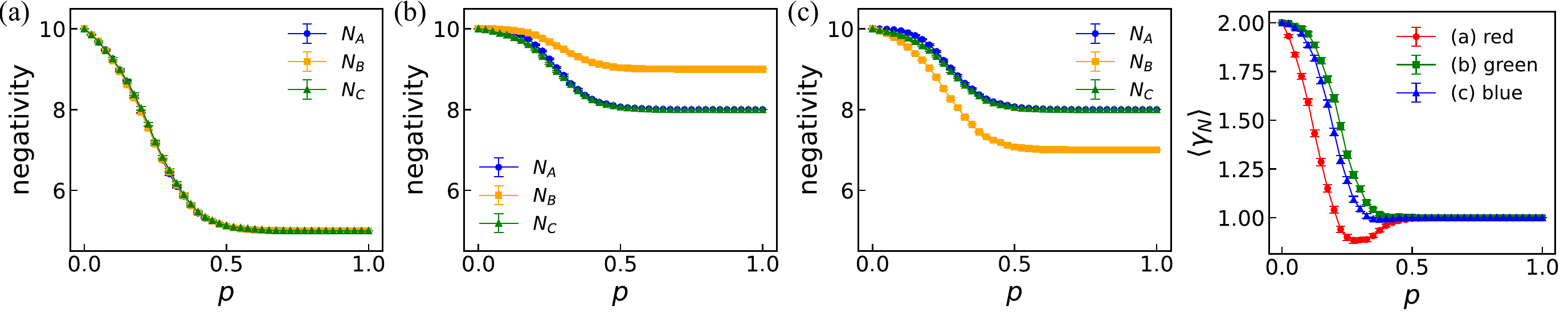}
\end{center}
\caption{Negativity and TEN for the systems shown in Fig.~\ref{Fig_TEN_partition2}.
Only the case (a) (left most) exhibits the location-independent value of the negativity.
However, the TEN (right most) exhibits the expected behavior from the color code to the TC in all the cases.
This result indicates the stability of the TEN, as a physical observable measuring topological entanglement.
Variance of the TEN has a similar behavior to that in Fig.~\ref{Fig_TEN}  in all three cases (not shown).
}
\label{FigNABC}
\end{figure*}

In Fig.~\ref{FigNABC}, we show the numerical data of the negativity and TEN concerning with
all the nine subsystems, labeled as $({\rm red}, A), ({\rm red},B), \cdots, ({\rm blue},C)$ in cases (a), (b) and (c),
where the first elements of the tuples refer to the color of the central plaquette of the complexes.
It is obvious that only the first case (a) with red-center plaquette has very close negativity for 
$A$, $B$, and $C$, whereas in the other two, the value of the negativity depends on $A$, $B$, and $C$.
However, the value of the TEN exhibits the expected behavior from the color code to the TC for 
all three cases. 
This implies that the subsystem combination of the TEN (as in Eq.~(\ref{def_TEN}) in the main text) excludes the system-size dependent part of the negativities even if each subsystem negativity depends on the shape of the subsystem.
This result indicates that the TEN is a stable physical quantity for measuring topological entanglement
of the mixed state.
As far as we know, this is the first observation of the suitability of the TEN as a measure of the evolution of 
the topological order.

\newpage
\bibliographystyle{apsrev4-2}
\bibliography{ref2}
\end{document}